\documentclass[unsortedaddress]{revtex4}

\usepackage{graphicx}
\usepackage{bm}

\begin{document}

\title{Numerical and Experimental Investigation of\\ 
Circulation in Short Cylinders}

\author{Akira Kageyama}
\email{kage@jamstec.go.jp}
\affiliation{
Earth Simulator Center, 
Japan Agency for Marine-Earth Science and Technology,
Yokohama 236-0001, Japan
}
\author{Hantao Ji}
\email{hji@pppl.gov}
\affiliation{
Princeton Plasma Physics Laboratory, 
Princeton, NJ 08543, USA
}
\author{Jeremy Goodman}
\email{jeremy@astro.princeton.edu}
\affiliation{
Princeton University Observatory, 
Princeton, NJ 08544, USA
}
\author{Fei Chen}
\email{fchen@pppl.gov}
\affiliation{
Princeton Plasma Physics Laboratory, 
Princeton, NJ 08543, USA 
}
\author{Ethan Shoshan}
\affiliation{
Rutgers University, Piscataway, NJ 08854, USA
}

\begin{abstract}
In preparation for an experimental study of magnetorotational
instability (MRI) in liquid metal, we explore Couette flows having
height comparable to the gap between cylinders, centrifugally stable
rotation, and high Reynolds number.  Experiments in water are compared
with numerical simulations.  Simulations show that
endcaps corotating with the outer cylinder drive a strong poloidal
circulation that redistributes angular momentum.  Predicted azimuthal
flow profiles agree well with experimental measurements.  Spin-down
times scale with Reynolds number as expected for laminar Ekman
circulation; extrapolation from two-dimensional simulations at $Re\le
3200$ agrees remarkably well with experiment at $Re\sim 10^6$.  This
suggests that turbulence does not dominate the effective viscosity.
Further detailed numerical studies reveal a strong radially inward
flow near both endcaps.  After turning vertically along the inner
cylinder, these flows converge at the midplane and depart the boundary
in a radial jet.  To minimize this circulation in
the MRI experiment, endcaps consisting of multiple, differentially
rotating rings are proposed.  Simulations predict that an adequate
approximation to the ideal Couette profile can be obtained with a few
rings.
\end{abstract}


\maketitle

\section{Introduction}
Laboratory experiments using liquid metal have made important tests of
magnetohydrodynamic (MHD) theories and processes.  Examples include
laboratory demonstration of the Alfv\'{e}n wave\cite{lundquist:1949a},
magnetofluid convection\cite{nakagawa:1955}, and dynamo
action\cite{gailitis:2000}.  Recently, a new kind of liquid-metal MHD
experiment, motivated by astrophysics, has been proposed by the
authors\cite{ji:2001,goodman:2002} to study
magnetorotational instability (MRI), which is believed to dominate the
transport of angular momentum in electrically conducting accretion
disks.  Liquid gallium will be used in a Couette flow between
cylinders of radii $r_1<r_2$ and angular velocities 
$0<\Omega_2<\Omega_1$ but $r_2^2\Omega_2>r_1^2\Omega_1$ so that the
flow will be stable against conventional Taylor-Couette instabilities
(TCI).

Both TCI and MRI 
are governed by the radial profile of azimuthal velocity, $v_\varphi(r)$.
In an inviscid fluid,
TCI occurs wherever the specific angular momentum, $J \equiv rv_\varphi$, 
decreases outwards, $dJ^2/dr<0$.
If the fluid is also a perfect electrical conductor,
MRI occurs when the angular velocity, $\Omega\equiv
v_\varphi/r$, decreases outwards, $d\Omega^2/dr<0$.  In this case, 
MRI can occur in an arbitrarily
weak axial magnetic  field; the field must at any rate
be weak enough so that the transit time of Alfv\'en waves
across the flow is less than $\Omega^{-1}$.
The usual situation in accretion disks, which are often excellent
conductors, is $\Omega\propto r^{-3/2}$ (Kepler's law)
so that TCI is stabilized but MRI is not.
In a fluid with large resistivity and finite but small viscosity, 
such as liquid metals, the range of flow parameters
unstable to MRI shrinks significantly, while the range unstable
to TCI is effectively unchanged. Therefore, laboratory flows
must be set up with precision in order to demonstrate MRI while suppressing
TCI. 

In previous linear stability analyses of gallium Couette
flow\cite{ji:2001,goodman:2002},
we adopted periodic boundary condition
in the vertical (axial) direction,
ignoring the effects of the top and bottom interior surfaces of the vessel
(``endcaps'').
The choice of vertical boundary conditions is probably inconsequential
when the height of the flow ($H$) is much larger than the
gap width, as in Taylor's classic experiments\cite{taylor:1923}.
Our experimental volume $\pi H (r_2^2-r_1^2)$
will be limited by the availability of gallium, a far
more expensive fluid than water, while the gap must be wide enough so
that the magnetic diffusion time is not much shorter than the rotation
period.  These considerations drive us to an
aspect ratio $H/(r_2-r_1)\sim O(1)$, in which the endcaps may assume great 
importance.

We have performed a water experiment
and complementary numerical simulations
to study the effects of the endcaps and, if possible,
to find a way to set up a short Couette flow that is unstable to MRI
yet stable against TCI.
Since the viscosities of the two fluids are similar,
standard visualization techniques in water serve to predict
the flow structure in the opaque liquid gallium,
at least in the absence of magnetic field.

Since the pioneering work by
Benjamin\cite{benjamin:1978a,benjamin:1978b,benjamin:1981}, 
TCI in
finite size cylinders have been studied
in detail with stationary outer
cylinders\cite{hall:1982,lucke:1984,aitta:1985,heinrichs:1986,
pfister:1988,tavener:1991,sobolik:2000,cliffe:1992,furukawa:2002,
mullin:2002,lopez:2003},
and with rotating outer cylinders\cite{schulz:2003}.
In our TCI-stable flows, the outer cylinder must
rotate.  To allow a wide gap, we use a relatively small radial aspect
ratio $\eta = r_1 / r_2 = 0.256$, so that the commonly used narrow-gap
approximation does not apply.  
The rotation rate of the inner
cylinder ($\Omega_1$) is so high that the Reynolds number
\begin{equation}\label{eq:Reynolds}
Re\equiv \frac{r_1 (r_2 - r_1) \Omega_1}{\nu} \sim O(10^6)\,,
\end{equation}
is orders of magnitude larger than typical Couette
flow experiments.  In connection with research on the tropospheric jet
stream, Dunst performed a water experiment in a short cylindrical
annulus with parameters similar to ours\cite{dunst:1972}.  We will
comment on this paper in section~\ref{section:summary}.

The outline of the paper is as follows.  Sec.\ref{section:experi}
describes the experimental apparatus and most of our experimental
results, except for spin-down measurements, which
are deferred to Sec.\ref{section:result}.
Numerical methods are described in
Sec.\ref{section:method}, and numerical simulations are presented and
compared with experiment in Sec.\ref{section:result}.  
Sec.\ref{section:summary} contains a summary of our main results and
a discussion of their significance.

\section{Water Experiment} \label{section:experi}
\subsection{Experimental Apparatus}
\begin{figure}\begin{center}
\includegraphics[width=0.5\textwidth]{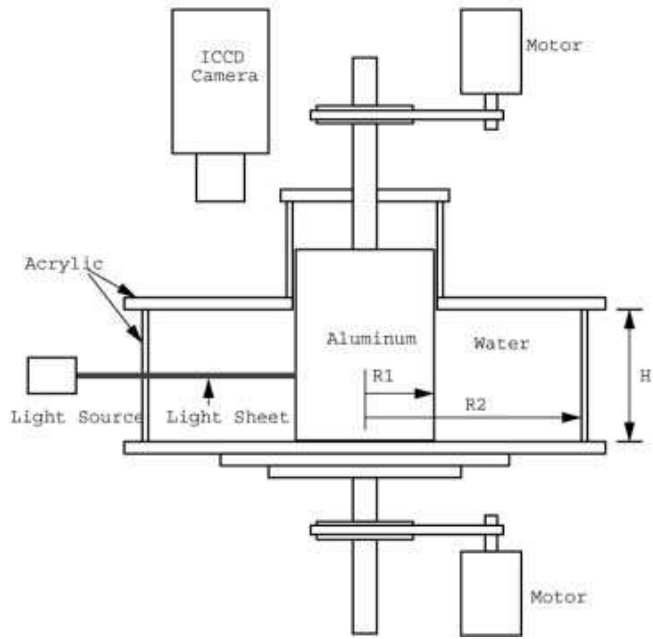}
\caption{\label{fig:apparatus}
Schematic of experimental apparatus. Radii of the inner and outer 
cylinders are $r_1$=3.8 cm and $r_2$=14.9 cm, respectively. The height 
is $H$=10 cm.}
\end{center}\end{figure}

The experiments are performed in a short circular Couette flow
illustrated in Fig.~\ref{fig:apparatus}. A container, made of
transparent acrylic plates and cylinders, is mounted 
on a stainless steel flange which is driven by a DC motor (1.25 HP, by 
Creative Industries). The speed is measured by a laser-based tachometer.
The inner cylinder, made of aluminum, is inserted through a lip seal 
from the top and is driven in the direction of the container 
by an AC motor (3 HP, by Lincoln Motors) with a controller. 
For given speeds of inner and outer cylinders, $\Omega_1$ and
$\Omega_2$, the flow typically requires
about 10-30 seconds to reach a steady state, in which
most of the flow measurements are performed.

\subsection{Measurement of Azimuthal Flow Profiles}

In order to measure flow profiles as a function of radius and
height, small particles with sizes on the order of 1 $\mu$m made of mica and
titanium dioxide (\lq\lq Sparkle\rq\rq by Lee S. McDonald, Inc.)
are mixed into the working fluid (water).  Being
small, the particles follow the flow well.
A sheet of light of approximate thickness of 0.5 cm is generated by a 
horizontal slot in front of a bright halogen light source.
An electronically gated intensified-CCD camera (by ITT Corp.)
images the flow illuminated by the sheet, whose height is adjustable.
Images are saved to a PC using a frame-grabber at a rate
of 60 images per second. Particles
appear in the images as streaks, whose length indicates
flow speed.
By combining measurements at many radii and heights, 
the azimuthal flow can be mapped out as a function of $r$ and $z$.
The measurements were calibrated, at various heights,
by imposing uniform rotation, $\Omega_1=\Omega_2=150$ rpm.

\begin{figure}\begin{center}
\includegraphics[width=0.5\textwidth]{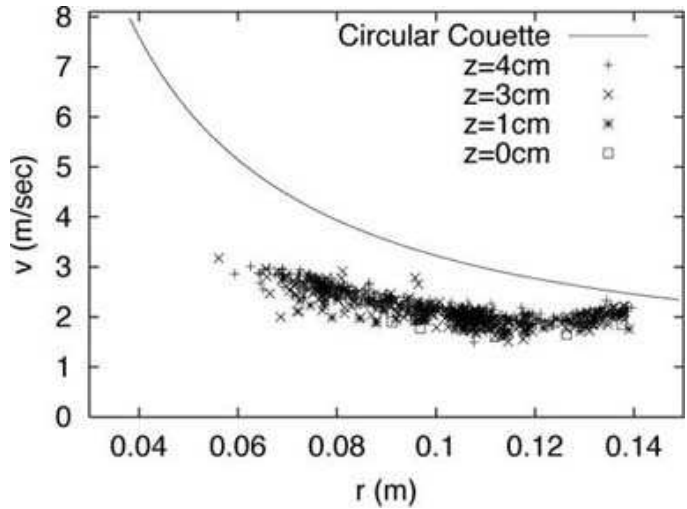}
\caption{\label{fig:vp_prof_experi}Measured $v_\varphi$ profile
at different $z$ when $\Omega_1$ = 2000 
rpm and $\Omega_2$= 150 rpm.}
\end{center}\end{figure}

The experimental results are shown in Fig.~\ref{fig:vp_prof_experi}
for the case of $\Omega_1$ = 2000 rpm and $\Omega_2$= 150 rpm.  There
are a few characteristics worth mentioning here.  First, the measured
velocity is significantly smaller than in an ideal, infinitely long
Couette flow having the same $\Omega_1$ and $\Omega_2$
[eq.(\ref{eq:couetteprof})].  The discrepancy is largest at small
radii. Secondly, the velocity must rise sharply from $\sim$ 3 m/s to
match the inner cylinder at $\sim$ 8 m/s.  Unfortunately, diagnostic
access to the flow is limited near the inner, outer, and top
boundaries.  Thirdly, the azimuthal velocity decreases with radius
except at the locations near the outer edge, whereas it would decrease
everywhere in the ideal Couette flow. 
Fourthly, the
dependence on $z$ is at most comparable to the experimental errors. This is
consistent with Taylor-Proudman theorem~\cite{batchelor:1967} which
predicts small $z$ variations in a rotating flow with small
viscosity.

The observed profile of azimuthal flow has unfavorable
implications for the proposed MRI experiments.
The goal is to set up
a flow unstable to MRI while stable to TCI. However,
the sharp decrease of $v_\varphi$ near the inner cylinder
will certainly incite TCI while the rest of the flow, because
$v_\varphi$ falls more slowly than intended,
will be more resistant to the MRI.
As a result, the system as a whole could have a mixture of both 
instabilities or, even worse, only the TCI.
The observed deviations from ideal Couette flow are due to the endcaps.
They need to be understood  and, if possible, to be minimized 
in order to demonstrate MRI unambiguously. We note that a proposed MRI experiment
using sodium~\cite{noguchi:2002} should suffer from the same complications 
since its aspect ratio is also small, \emph{viz.} $H/(r_2-r_1)=2$.

\section{Numerical Methods} \label{section:method}
\subsection{Mathematical Model and Algorithm}
\begin{figure}\begin{center}
\includegraphics[width=0.9\textwidth]{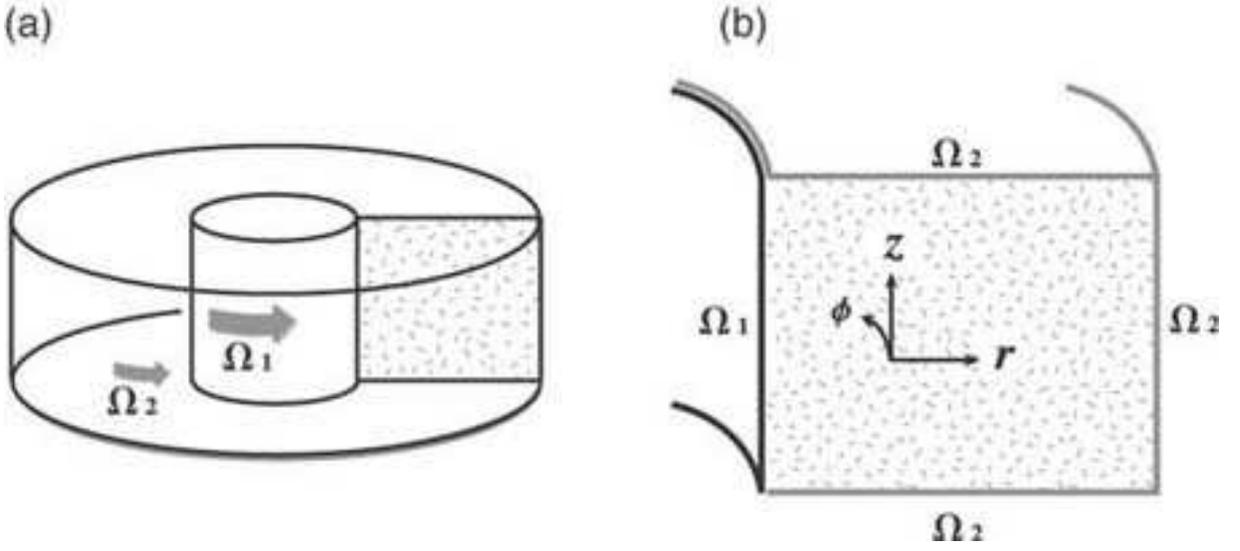}
\caption{\label{fig:system}Illustrations of physical system (a) 
and simulation system (b).}
\end{center}\end{figure}
We have developed a 2-dimensional computer code to simulate
the water experiment described in the previous section.
We use the stream function-vorticity method\cite{ferziger:2002}
in cylindrical coordinates ($r, \varphi, z$), as illustrated
in Fig.~\ref{fig:system}.
Assuming axisymmetry, $\partial_\varphi(v_r,v_\varphi,v_z) = 0$,
and incompressibility, $\nabla \cdot \bm{v} = 0$,
we can express the poloidal components of the flow in terms of
a stream function, $\psi$:
\begin{eqnarray}
  v_r & = &  \frac{1}{r} \frac{\partial \psi}{\partial z}, \label{eq:sim_01}\\*
  v_z & = &  - \frac{1}{r} \frac{\partial \psi}{\partial r}. \label{eq:sim_02}
\end{eqnarray}
The azimuthal component of vorticity becomes
\begin{equation}
  \left( \frac{1}{r} \frac{\partial^2}{\partial r^2}
    - \frac{1}{r^2} \frac{\partial}{\partial r}
    + \frac{1}{r} \frac{\partial^2}{\partial z^2} \right) \psi
   = \omega_\varphi.     \label{eq:sim_03}
\end{equation}
Since the normal component of velocity vanishes,
the boundary conditions for $\psi$ are
\begin{eqnarray}
  \psi & = & 0 \qquad \text{at} \qquad  r=r_1, r_2, \label{eq:sim_09a}\\
  \psi & = & 0 \qquad \text{at} \qquad z= \pm H/2.\label{eq:sim_09}
\end{eqnarray}

The boundary conditions for $v_\varphi$ are as follows.
At the inner and outer cylindrical walls,
\begin{eqnarray}
  v_\varphi &=& r_1 \Omega_1 \qquad \text{at} \qquad  r=r_1.\label{eq:sim_10} \\
  v_\varphi &=& r_2 \Omega_2 \qquad \text{at} \qquad r=r_2. \label{eq:sim_11}
\end{eqnarray}
Since the top and the bottom endcaps of the vessel
rotate with the outer cylinder in our apparatus,
the boundary condition there are
\begin{equation}
  v_\varphi = \frac{r}{r_2} \Omega_2 \qquad 
                 \text{at} \qquad z=\pm H/2. \label{eq:sim_12}
\end{equation}

The no-slip conditions on $v_r$ at $z=\pm H/2$ and on $v_z$ at $r=r_1,r_2$
yield boundary conditions for $\omega_\varphi$ \emph{via} 
eqs.(\ref{eq:sim_01})--(\ref{eq:sim_03}):
\begin{eqnarray}
  \omega_\varphi & = & \frac{1}{r} \frac{\partial^2 \psi}{\partial r^2}
                       \ \ \ \ \ \text{at}\ \ r = r_1, r_2, \label{eq:sim_14} \\*
  \omega_\varphi & = & \frac{1}{r} \frac{\partial^2 \psi}{\partial z^2}
                       \ \ \ \ \ \text{at}\ \ z = \pm H/2. \label{eq:sim_15}
\end{eqnarray}

The fundamental variables in our numerical simulation are 
$v_\varphi$ and $\omega_\varphi$.  Their governing equations are
\begin{eqnarray}
  \frac{\partial \omega_\varphi}{\partial t} & = &
     \frac{\partial D_r}{\partial z} - \frac{\partial D_z}{\partial r}
   + \nu \left( \nabla^2 - \frac{1}{r^2} \right) \omega_\varphi,\label{eq:sim_04}\\*
  \frac{\partial v_\varphi}{\partial t} & = &
     - \left( v_r \frac{\partial v_\varphi}{\partial r}
            + v_z \frac{\partial v_\varphi}{\partial z}
            + \frac{v_r v_\varphi}{r} \right)
     + \nu \left( \nabla^2 - \frac{1}{r^2} \right) v_\varphi, \label{eq:sim_05}
\end{eqnarray}
where
\begin{eqnarray}
  D_r & = & \frac{J}{r^2} \left( \frac{\partial J}{\partial r} \right)
          + \frac{\omega_\varphi}{r} \left( \frac{\partial \psi}{\partial r}
          \right), \label{eq:sim_06}                     \\*
  D_z & = & \frac{\omega_\varphi}{r} \left( \frac{\partial \psi}{\partial z} \right)
          + \frac{J}{r^2} \left( \frac{\partial J}{\partial z}
          \right), \label{eq:sim_07}                     \\*
    J & = & r v_\varphi,  \label{eq:sim_13}
\end{eqnarray}
and
\begin{equation}
  \nabla^2 = \frac{\partial^2}{\partial r^2}
         + \frac{1}{r} \frac{\partial}{\partial r}
         + \frac{\partial^2}{\partial z^2}. \label{eq:sim_08}
\end{equation}

We use second-order spatial differences on a uniform grid, with
typical size $N_r\times N_z = 100 \times 100$,
and a fourth-order Runge-Kutta method for the temporal integration,
with typical time step $\delta t = 7.273 \times 10^{-3}(r_2-r_1)/r_1\Omega_1$.

The algorithm for each time step is as follows.
\begin{enumerate}
\item Integrate the basic eqs.(\ref{eq:sim_04}) and (\ref{eq:sim_05})
      to get new $\omega_\varphi(i,k)$ and $v_\varphi(i,k)$
      in the bulk region ($2 \le i \le N_r-1$, $2 \le k \le N_z-1$).
\item Solve eq.(\ref{eq:sim_03}) for $\psi$ with
      the new $\omega_\varphi$ as source term and boundary conditions
 (\ref{eq:sim_09a})--(\ref{eq:sim_09}).
\item Set the boundary values of $\omega_\varphi$ and $v_\varphi$
      using eqs.(\ref{eq:sim_10}), (\ref{eq:sim_11}),
      (\ref{eq:sim_14}) and (\ref{eq:sim_15}).
\item Get the auxiliary variables $v_r$, $v_z$, $D_r$, $D_z$, and $J$ 
      from eqs.(\ref{eq:sim_01})-(\ref{eq:sim_02}) \& (\ref{eq:sim_06})--(\ref{eq:sim_13}).
\end{enumerate}

\subsection{Simulation Method}

The boundary conditions (\ref{eq:sim_10}), (\ref{eq:sim_11}), and
(\ref{eq:sim_12}) require a jump of $v_\varphi(r)$ at the corners
$(r=r_1,z=\pm H/2)$.  A commonly used technique to avoid this
singularity is to make a small ``buffer region'' $r_1\le r\le r_1 +
\epsilon$, in which $v(r,\pm H/2)$ varies smoothly\cite{mullin:2000}.
But we found that such a ``buffer region'' is not necessary in our
scheme.  Following the standard technique, we first set $\epsilon$ to
$10\%$ of the gap width $r_2-r_1$, and then gradually reduced it to
the radial mesh size, $\Delta r$.  In the latter case the angular
velocity of the boundary jumps from $\Omega_1$ to $\Omega_2$ between
the innermost two grid points ($i=1,2$).  We confirmed that simulation
results are not affected by the size of $\epsilon$.  All calculations
shown in this paper use $\epsilon = \Delta r$.

All calculations begin with both
the boundaries and the fluid at rest.
The boundaries (side walls and endcaps) accelerate to
their final angular velocities over a short time interval $0\le t\le\tau$.
We confirmed through many trials that the final state of the flow
does not depend upon $\tau$.
All simulations shown here were calculated with 
$\tau=(r_1-r_2)/(r_1\Omega_1)$, our unit of time.

\section{Results of Simulations} 
\label{section:result}

\subsection{Parameters and units}
The simulations shown in this section use the same
dimensions and rotation rates as
as the experiments described in Sec.~\ref{section:experi}.
The computational units of length, velocity, and time are
$\ell=r_2-r_1$, $v=r_1\Omega_1$, and $\ell/v$.

The Reynolds number of the experiment [eq.(\ref{eq:Reynolds})] is too
high for direct numerical simulation.  So we start by simulating
very low $Re$ and increase it until we find characteristics
of the flow that depend only weakly on $Re$ or follow a clear scaling.
The largest simulated $Re$ that we report is $3200$.

\subsection{Low $Re$ flows}

\begin{figure}\begin{center}
\includegraphics[width=0.5\textwidth]{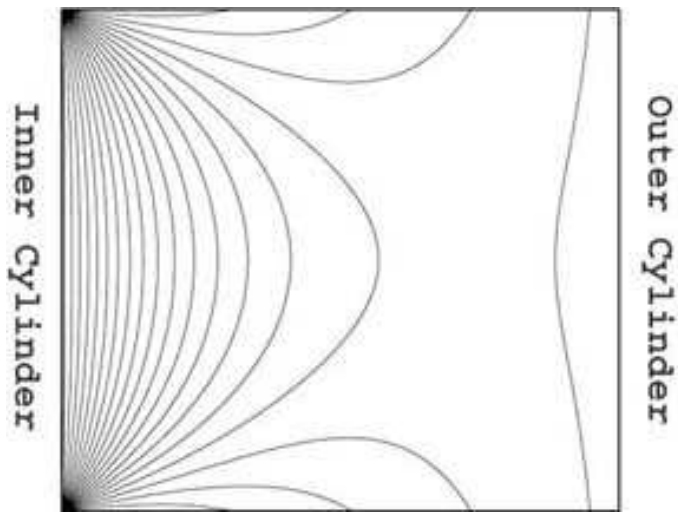}
\caption{\label{fig:vp_contour_Re=1_with_label}Simulated 
$v_\varphi$ profile when
$Re=1$.}
\end{center}\end{figure}

The sudden acceleration of the boundaries at the beginning
of the simulation induces an intense flow inside the vessel.
At small values of the final rotation rate, \emph{i.e.} small $Re$,
the fluid quickly relaxes to a laminar steady state.

Figure \ref{fig:vp_contour_Re=1_with_label} shows the azimuthal speed,
$v_\varphi$, for $Re=1$.
The maximum speed is achieved at the inner cylinder
(on left in Fig.~\ref{fig:vp_contour_Re=1_with_label}),
where  $v_\varphi=1$.
In this low $Re$ limit, poloidal flow is almost absent; the
maximum value of $v_r$ and $v_z$ is $7.5\times 10^{-4}$.
Note that the jump in $\Omega$ at the lefthand corners of the boundary
does not prevent a smooth solution elsewhere.

Actually, as $Re\to0$,
the solution for the flow can be found in closed form.
If we set $v_r=v_z=\omega_\varphi=0$ in the basic equations
(\ref{eq:sim_04})--(\ref{eq:sim_05}),
the stationary azimuthal flow $v_\varphi$ satisfies
\begin{equation}
   \nu \left( \nabla^2 - \frac{1}{r^2} \right) v_\varphi
   = 0,     \label{eq:sim_23}
\end{equation}
with the boundary conditions (\ref{eq:sim_10})--(\ref{eq:sim_12}).
Wendl\cite{wendl:1999} has given the analytical solution of
this equation with slightly different boundary conditions,
corresponding to $\Omega_2=0$.
The solution to our problem is given simply by adding
a uniformly rotating component $r \Omega_2$ to Wendl's solution.
The $v_\varphi$ profile shown in 
Fig.~\ref{fig:vp_contour_Re=1_with_label} is essentially
identical to the analytical solution thus constructed.
This serves as one benchmark for our code.

\subsection{High $Re$ flows}
\begin{figure}\begin{center}
\includegraphics[height=0.9\textheight]{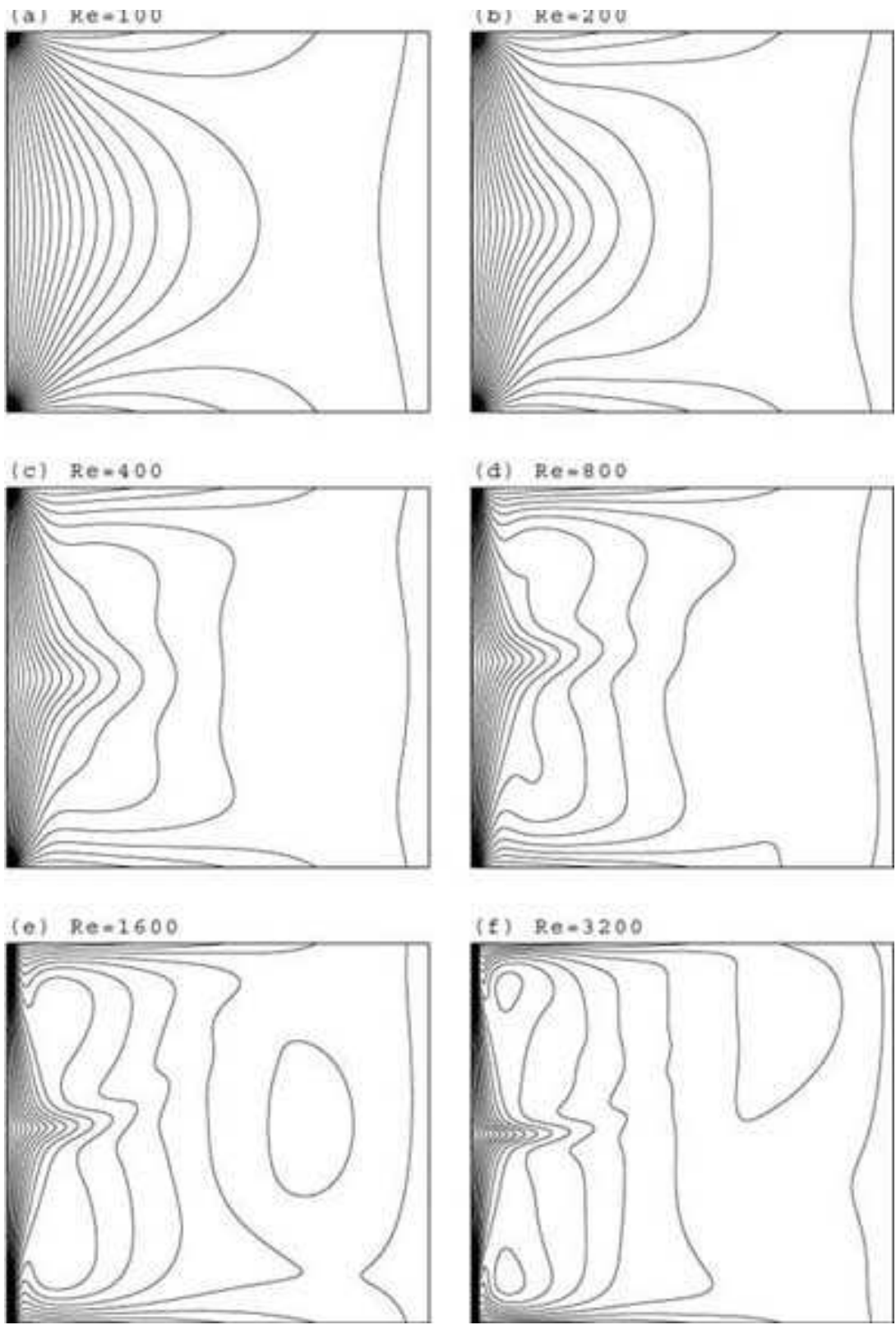}
\caption{\label{fig:vertical_combined_vp}$V_\varphi$ profile.}
\end{center}\end{figure}

We have seen that poloidal flow is negligible in the low $Re$ regime.
As we increase $Re$, poloidal circulation develops and the azimuthal
flow changes.  Figure~\ref{fig:vertical_combined_vp} shows cross
sections in the poloidal ($r,z$) plane for different $Re$ numbers, from
$100$ to $3200$.  Each panel in Fig.~\ref{fig:vertical_combined_vp} is
a snapshot of the nonlinearly saturated state.

Figure~\ref{fig:vertical_combined_vp}(a) shows that $v_\varphi$ has much
the same pattern at $Re=100$ as at $Re=1$
(Fig.~\ref{fig:vp_contour_Re=1_with_label}).  In both cases, it is
symmetric about the horizontal plane $z=0$ and time independent.  The
flow becomes unsteady at $Re>400$.  The asymmetric profile of $Re=800$
[Fig.~\ref{fig:vertical_combined_vp}(d)] results from unsteady flow.
All flows above this Reynolds number fluctuate, with an amplitude that
increases with $Re$.

One of the important features shown in
Fig.~\ref{fig:vertical_combined_vp} is that the contours of
$v_\varphi$ tend to be parallel to the rotation axis ($z$).  This is a
manifestation of the Taylor-Proudman theorem, \emph{viz.}, that
low-frequency horizontal motions tend to be independent of height in
an inviscid fluid rotating about a vertical
axis\cite{chandrasekhar:1968}.  The Taylor-Proudman theorem
is usually discussed for an almost rigidly rotating fluid,
but as shown in the Appendix, a similar tendency exists in
differentially rotating flow provided $dJ^2/dr>0$.

Another characteristic feature of the $v_\varphi$ contours in
Fig.~\ref{fig:vertical_combined_vp} is their tendency to concentrate
towards the inner cylinder (at left) at large $Re$.  This shows the
development of a boundary layer.  Boundary layers also develop on
the top and bottom endcaps [see panels (e) and (f)].  Note also the
increasingly sharp protrusion of the contours on the inner cylinder
near the midplane.

\begin{figure}\begin{center}
\includegraphics[height=0.9\textheight]{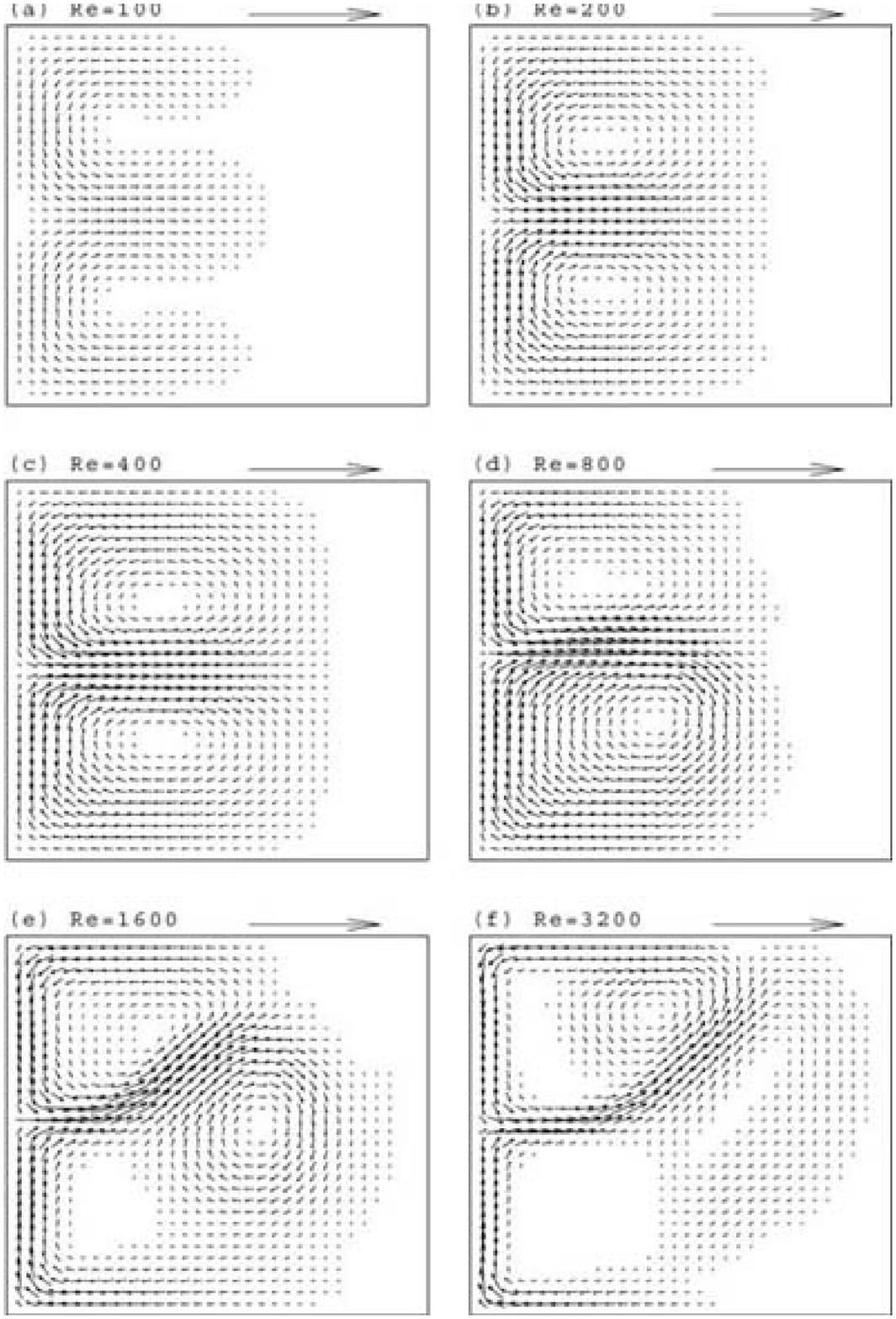}
\caption{\label{fig:vertical_combined_vrz}Poloidal flow profile.}
\end{center}\end{figure}

The development of the 
poloidal ($v_r,v_z$) flow is equally
interesting.  It is shown in the vector plots of
Fig.~\ref{fig:vertical_combined_vrz}
for the same simulations as in
Fig.~\ref{fig:vertical_combined_vp}.
Vectors appear at every third grid point in $r$ and $z$.
Their lengths indicate that the
poloidal component becomes stronger at higher $Re$.
The arrow lengths are linearly normalized.
The amplitude of the unit velocity, 
that is the rotation speed of the inner cylinder,
is indicated by arrows on the top of the panels (a) to~(f).
A radially outward, jet-like flow
is seen near the midplane in the higher-$Re$ simulations.
It is undoubtedly the counterpart of the spike in
the $v_\varphi$ contours in Fig.~\ref{fig:vertical_combined_vp}.
This jet-like flow was not expected before we began
our numerical simulations.
Its structure will be analyzed later.
Here we show the temporal behavior of the poloidal flow, 
including the jet.

\begin{figure}\begin{center}
\includegraphics[width=0.9\textwidth]{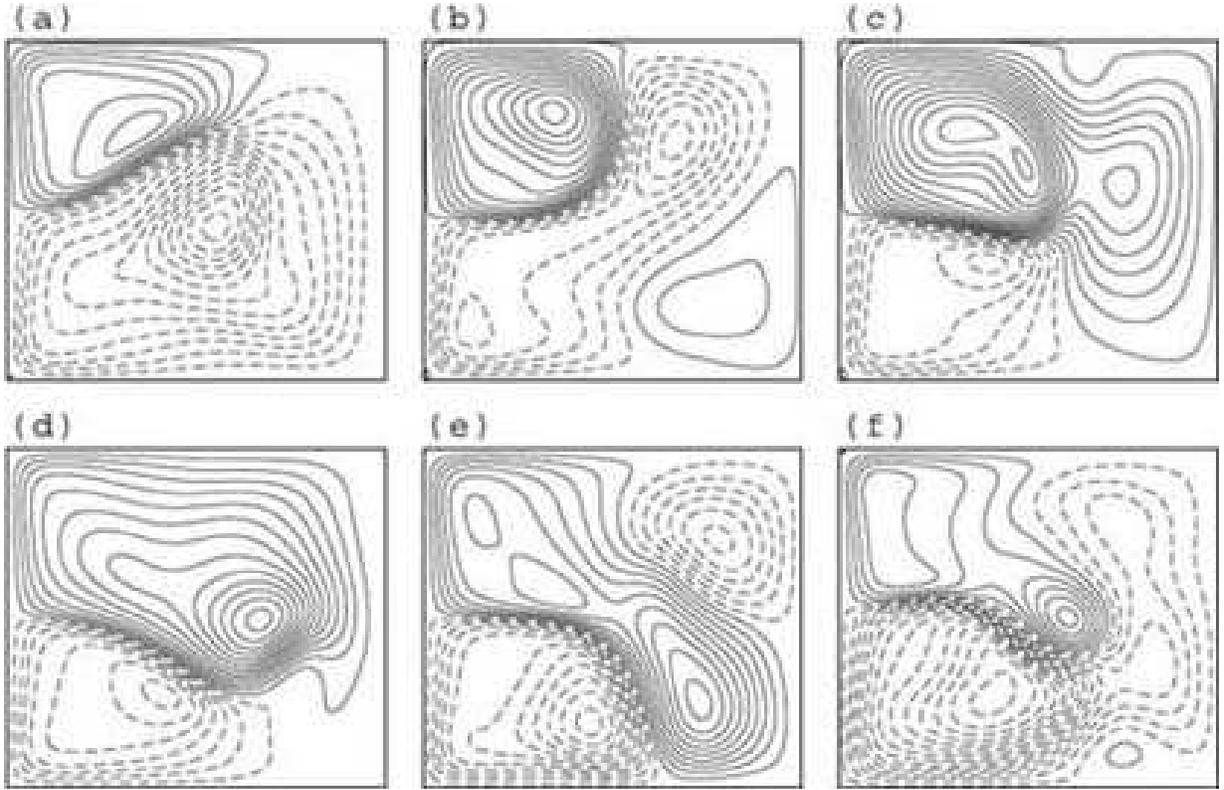}
\caption{\label{fig:jet_time_develop}Contour plots
of the stream function in a time sequence showing the
oscillatory motion of jet ($Re=3200$).}
\end{center}\end{figure}

The jet flaps unsteadily.  Figure~\ref{fig:jet_time_develop} shows the
time development of the stream function, $\psi$, for $Re=3200$ over
almost half of the period of oscillation of the jet.
The period of the oscillation is about $36.95$ in the normalized time.
Figure~\ref{fig:jet_time_develop}(a) is a snapshot taken at $t = 1326.68$,
and subsequent panels (b) to (f) are taken with regular $5.82$ interval time.
Throughout the oscillation, the
root of the jet (at $r=r_1, z=0$) remains fixed while its tip flaps
violently up and down.  The vector plot in panel (f) of
Fig.~\ref{fig:vertical_combined_vrz} shows a snapshot of this flapping
motion at $t=1294.67$.

\subsection{Profile of Azimuthal Flow}

\begin{figure}\begin{center}
\includegraphics[width=0.5\textwidth]{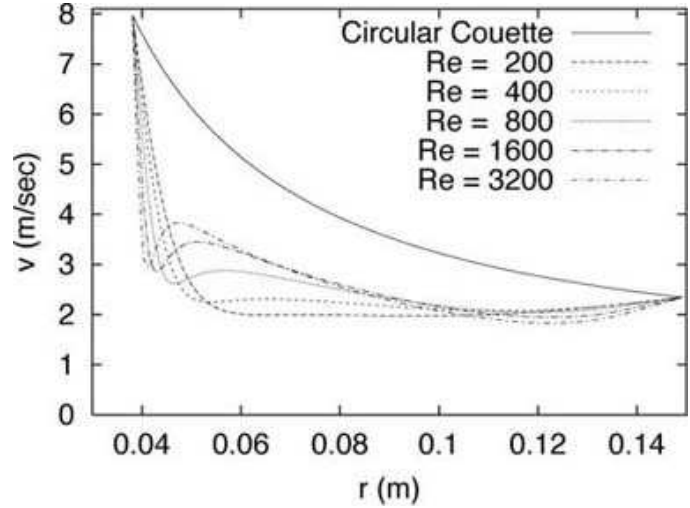}
\caption{\label{fig:vp_r0200_to_r3200}Simulated $v_\varphi$ profile
at $z=0$.}
\end{center}\end{figure}

\begin{figure}\begin{center}
\includegraphics[width=0.5\textwidth]{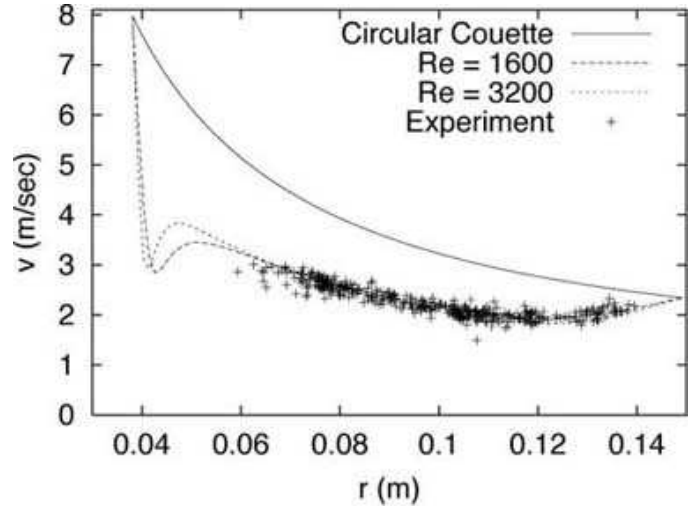}
\caption{\label{fig:vp_sim_and_exp}Comparison of $v_\varphi$
profiles obtained by simulation and experiment.}
\end{center}\end{figure}

\begin{figure}\begin{center}
\includegraphics[width=0.5\textwidth]{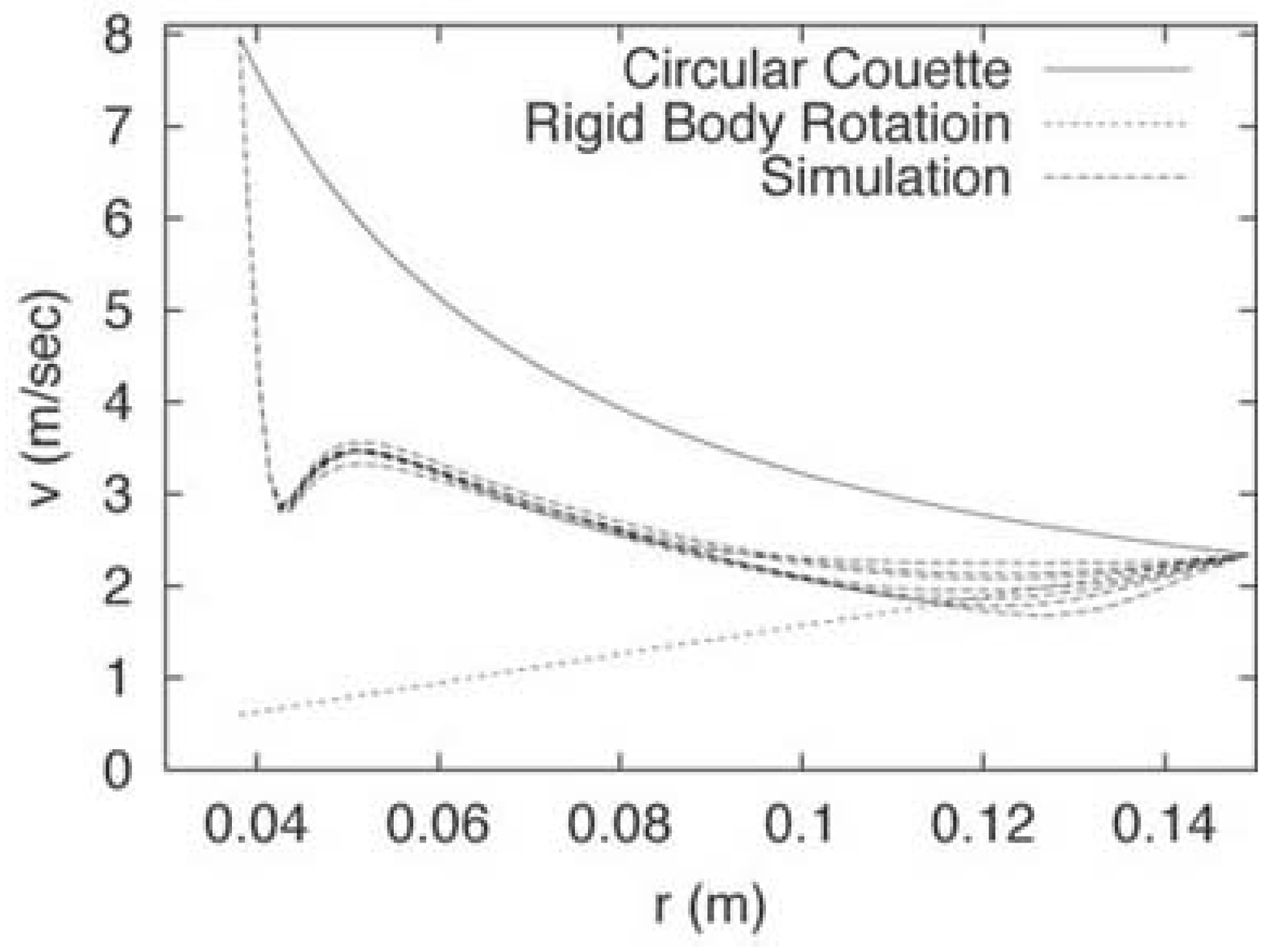}
\caption{\label{fig:vp_r1600_series}Time variation of
$v_\varphi$ profile for $Re=1600$.
The rigid rotation profile of the endcaps and outer cylinder shown
for comparison.}
\end{center}\end{figure}

Figure~\ref{fig:vp_r0200_to_r3200} shows $v_\varphi(r)$
for different Reynolds numbers ($200\le Re\le3200$)
at $z=4\text{cm}$.
(This is the height where the most reliable experimental data can be obtained.)
One can see from this figure that 
the $v_\varphi$ profiles are consistent for the higher $Re$ flows,
$Re=800$, $1600$, and $1600$.
Note also that the profiles for $Re=1600$ and $3200$ are very close, 
suggesting possible convergence at large $Re$.

In Fig.~\ref{fig:vp_sim_and_exp},
we have superimposed the experimental data for $v_\varphi$
at $z=4\text{cm}$ (shown also in Fig.~\ref{fig:vp_prof_experi})
on the corresponding profiles from simulations
at $Re=1600$ and $3200$.
The agreement is remarkable when one considers that
the highest $Re$ achieved in simulations, which are in 2D, 
is a factor of 300 smaller than that of the experiment,
which is 3D.

We have seen in Fig.~\ref{fig:vertical_combined_vp} that 
the high $Re$ ($>400$) flows are time dependent.
To show the amplitude of the temporal fluctuations,
snapshots of $v_\varphi$ for $Re=1600$
are superimposed in Fig.~\ref{fig:vp_r1600_series}.
(The curves in Fig.~\ref{fig:vp_sim_and_exp} are time averages.)
We can see that the amplitude is
larger in the outer half of the flow ($r>9\text{cm}$).
This can be explained by the flapping motion of the jet
(see Fig.~\ref{fig:jet_time_develop}).

Both experiment and simulation indicate that
the $v_\varphi$ profile is concave in the outer half of the flow
(Fig.~\ref{fig:vp_sim_and_exp}), and that
$v_\varphi(r)$ is an increasing function of radius near 
the outer cylinder ($12\text{cm} < r < 15\text{cm}$).
The fluid in this region rotates almost
rigidly, on average, at the angular velocity of the outer
cylinder and endcaps ($\Omega_2$).

\begin{figure}\begin{center}
\includegraphics[width=0.5\textwidth]{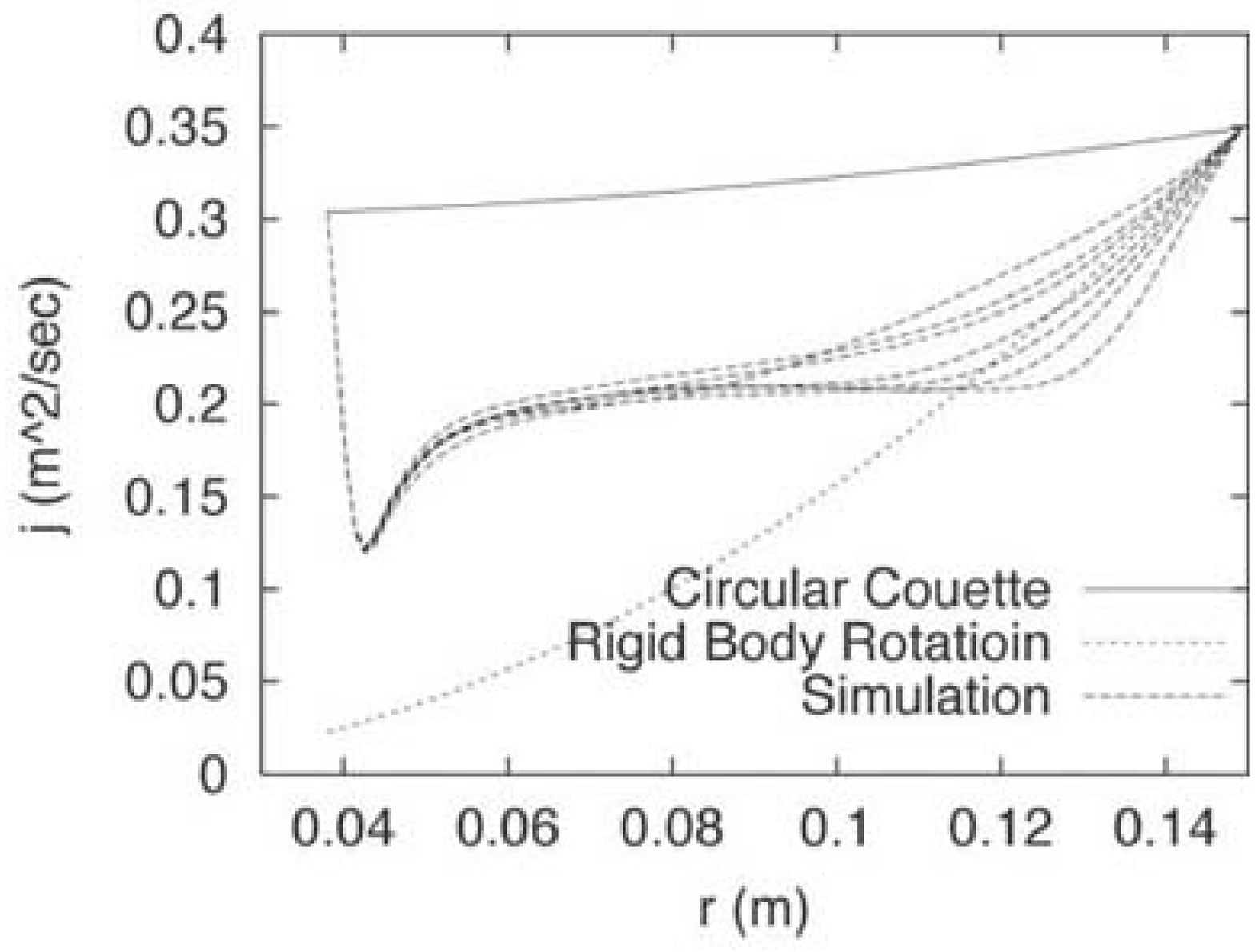}
\caption{\label{fig:jp_r1600_series}Time variation of
angular momentum $J=r v_\varphi$ for $Re=1600$.}
\end{center}\end{figure}

Snapshots of the angular momentum $J=r v_\varphi$
are shown in Fig.~\ref{fig:jp_r1600_series} for the same
simulation and as in Fig.~\ref{fig:vp_r1600_series}.
Note that $dJ/dr >0$ for the ideal Couette flow (solid curve)
since we aim for stability against
the Taylor-Couette mode.
An interesting feature of this figure is that the simulated
$J$ curve is even flatter than the ideal profile.
This is a consequence of the poloidal circulation and jet, which
tend to mix angular momentum in the interior of the flow.

\subsection{Jet}

\begin{figure}\begin{center}
\includegraphics[width=0.8\textwidth]{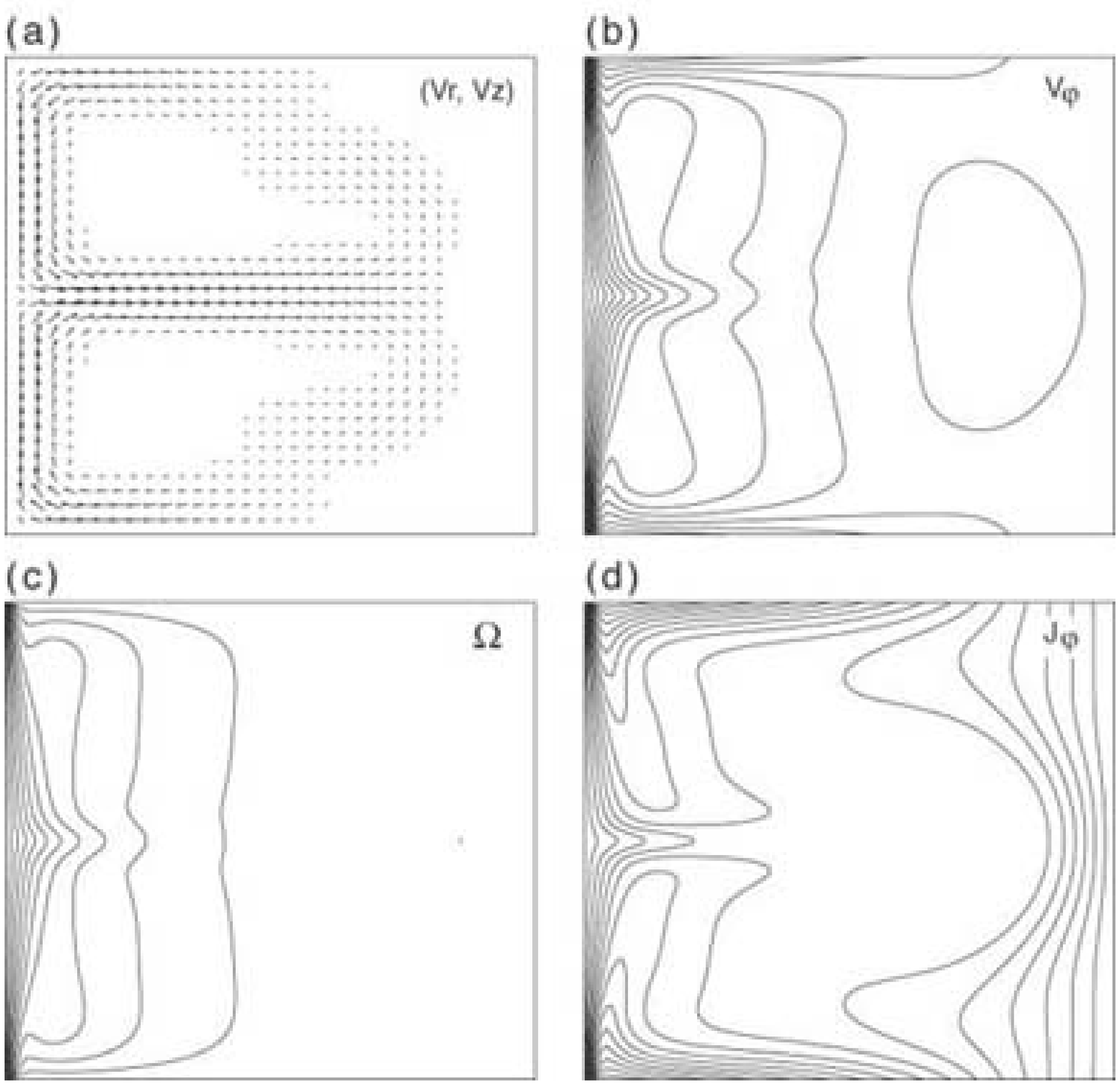}
\caption{\label{fig:vertical_sym_r1600}Flow under enforced symmetry 
at $Re=1600$: (a) poloidal flow;
(b) contours of azimuthal velocity $v_\varphi$; 
(c) angular velocity $\Omega$; and (d) angular velocity $J$.}
\end{center}\end{figure}

In order to extract essential features of the
spatial structure of the flow,
it would be convenient if we could suppress its temporal fluctuations,
especially the flapping of the jet.
For this purpose, we have made a set of simulations with
the following symmetries imposed:
\begin{eqnarray*}
  v_r(r,z)       & = & v_r(r,-z),       \\*
  v_\varphi(r,z) & = & v_\varphi(r,-z), \\*
  v_z(r,z)       & = & -v_z(r,-z),
\end{eqnarray*}
Fig.~\ref{fig:vertical_sym_r1600} shows the result when $Re=1600$.
One can perhaps interpret this state as an average
of the flow over one period of the jet's flapping motion.

The panel (a) in Fig.~\ref{fig:vertical_sym_r1600} clearly shows
the structure of the poloidal circulation.
It consists of three main parts;
(i) inward flow in boundary layers at the endcaps;
(ii) axial flow towards the midplane on the inner cylinder;
and
(iii) an outward jet centered on the midplane.

The boundary layer on the inner cylinder has a characteristic
triangular shape [panel (b) in Fig.~\ref{fig:vertical_sym_r1600}].
The contours are ``squeezed'' by the inward boundary-layer flows near
the top and the bottom caps, while, the tip of the triangle is
``pulled'' by the outward jet flow.

The contour lines of $v_\varphi$ and $\Omega$
[Fig.~\ref{fig:vertical_sym_r1600}(b) and (c)] suggest that the flow
attempts to obey the Taylor-Proudman theorem outside of the boundary
layers and the jet.  Since the latter regions are thin, especially at
high $Re$, viscous forces becomes important there and the
Taylor-Proudman theorem does not hold.

The contour lines of $\Omega$,
Fig.~\ref{fig:vertical_sym_r1600}(c), show that
the the outer half of the fluid rotates approximately at the
angular velocity of the outer cylinder and endcaps,
as we have already seen in 
Figs.~\ref{fig:vp_r1600_series} and~\ref{fig:jp_r1600_series}.
We have also seen that
the angular momentum $J$ tends to be uniform in the interior
due to the poloidal circulation, as is
clearly shown by the central void in the contour
plot of $J$ in Fig.~\ref{fig:vertical_sym_r1600}(d).

\begin{figure}\begin{center}
\includegraphics[width=0.45\textwidth]{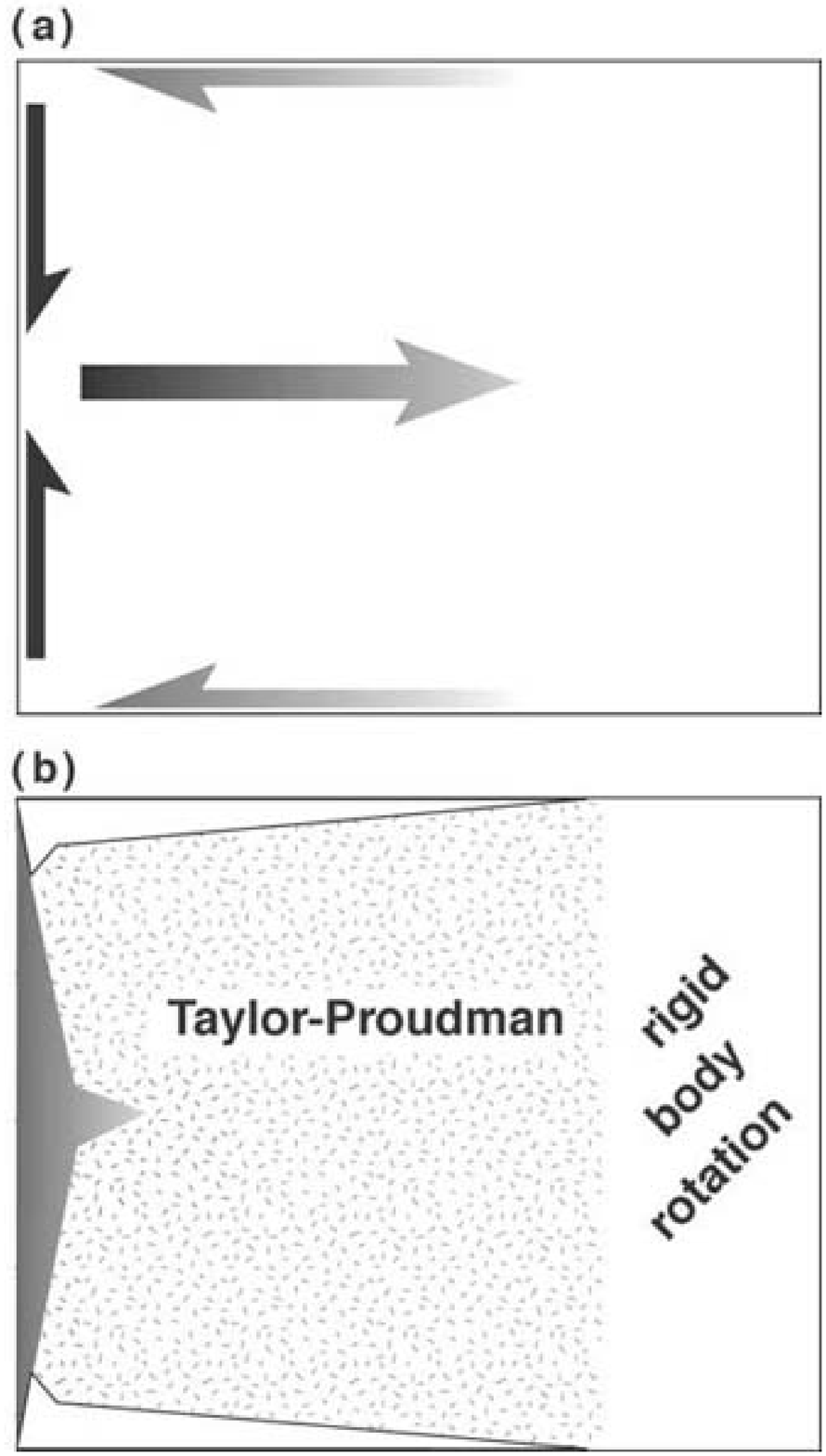}
\caption{\label{fig:schematic_flow_structure}Schematic flow
structure. (a) The Poloidal flow in the boundary layers
and the jet. (b) Azimuthal flow structure.}
\end{center}\end{figure}

The structures of the azimuthal and poloidal flows are
summarized schematically in
Fig.~\ref{fig:schematic_flow_structure}.

One interesting but unexpected finding of these simulations
is the existence of the jet.
Its characteristics are summarized as follows:
(1) The jet becomes thinner with increasing $Re$, and its width is
similar to that of the boundary layers on the caps;
and (2) the jet is steady and symmetric at low $Re$ but flaps
vertically above a critical Reynolds number between $400$ and $800$.

Experimentally, it is not straightforward to confirm the existence of
a jet-like radial flow directly. Since the maximum radial velocity in
the simulation is only a few percent of the large azimuthal flow, the
streaks on the camera images would rotate only a few degrees, too
little to be resolved by our measurements.  At lower rotation speeds,
we were able visually to follow relatively large and neutrally buoyant
particles in the water. We observed rather rapid outward motions at
the midplane after the particles were ``sucked'' into the boundary
layers at the top or bottom endcaps.  This is consistent with the
jet-like flows indicated by the simulations.

\subsection{Boundary Layers}
The width of the boundary layers (including the jet)
depends on $Re$.
As Fig.~\ref{fig:vertical_combined_vrz} indicates,
higher $Re$ causes thinner boundary layers.
Actually, the inward (negative $v_r$) flow on the top and bottom boundaries
is induced by the same mechanism as classical Ekman circulation:
a viscous reduction of $v_\varphi$ in the boundary layer, leading to
an imbalance between outward centrifugal
force and the pressure gradient.
The boundary layers at the endcaps in our system do not have uniform width.
As Figs.~\ref{fig:vertical_sym_r1600}(a),~(b), and~(d) indicate,
these layers are thick near the inner cylinder, reaching
roughly $10\%$ of the vessel's height ($H$).
The width monotonically declines with increasing $r$,
disappearing into the rigid rotation part of the outer part of the fluid.
[See also Fig.~\ref{fig:schematic_flow_structure}(b).]


For small departures from rigid rotation, the Ekman layer thickness
is $\delta_E=\sqrt{\nu/\Omega}$, where $\nu$ is the kinematic viscosity.
Our system is very far from rigid rotation, so it is not immediately
clear what to substitute for $\Omega$.  If one uses the mean frequency
$\sqrt{\Omega_1\Omega_2}$, then for our geometry, $\delta_E/H\approx
1.24 Re^{-1/2}$, hence $\approx 3\%$ at $Re=1600$.  In fact, we
estimate from our simulations that the fractional thickness of the
boundary layers is $\sim 10\%$ at this Reynolds number.  For small
departures from a differentially rotating state, however, we believe
that it is more appropriate to scale $\delta_E$ with the epicyclic
frequency, 
\begin{equation}\label{epicyclic} \kappa=
\left(\frac{1}{r^3}\frac{\partial J^2}{\partial r}\right)^{1/2}.
\end{equation}
This is the maximum frequency of small axisymmetric
motions (inertial oscillations) in the inviscid interior of the fluid,
so it represents the inertial forces that must be balanced by viscous
ones in order to drive a radial flow along the boundaries.  Since
$\kappa$ reduces to $2\Omega$ for rigid rotation, we take
$\delta_E=\sqrt{2\nu/\bar \kappa}$.  A characteristic value for $\kappa$ is
\begin{equation}\label{eq:kappabar}
\bar\kappa = 2\left(\frac{r_2^4\Omega_2^2-r_1^4\Omega_1^2}
{r_2^4-r_1^4}\right)^{1/2}
\end{equation}
This leads to $\delta_E/H= 3.39 Re^{-1/2}$, or $\sim 8.5\%$ at
$Re=1600$, which is about three times larger than the previous
estimate and closer to the results of the simulations.

A prediction of the latter scaling is that the Ekman-layer thickness
should increase along a sequence in which
$(r_2^2\Omega_2)/(r_1^2\Omega_1)$ approaches unity (from above)
while the mean rotation $\sqrt{\Omega_2\Omega_1}$ is constant.

\subsection{Flow in Shorter Cylinder}

\begin{figure}\begin{center}
\includegraphics[width=0.45\textwidth]{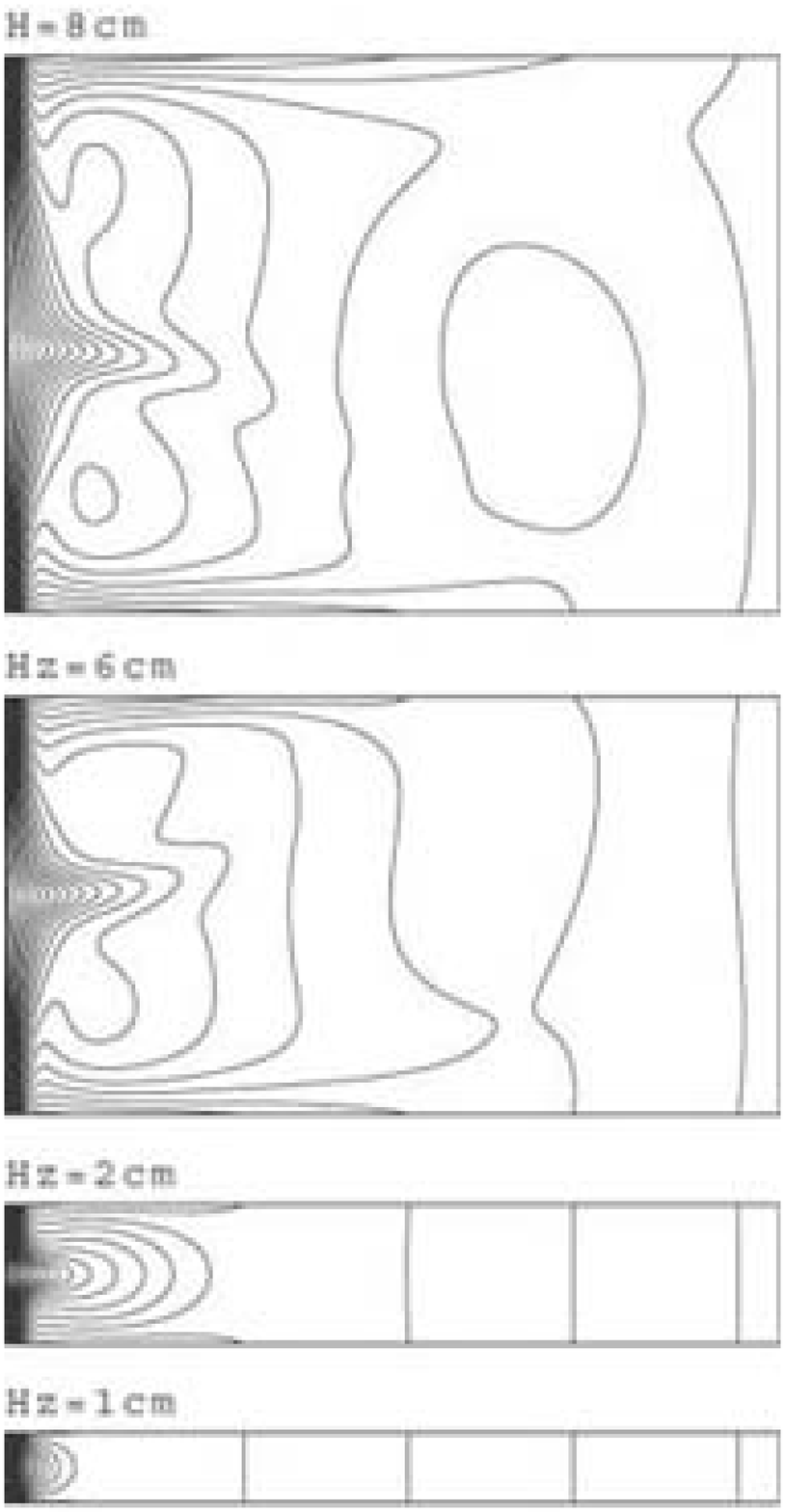}
\caption{\label{fig:short_heights}Profiles of azimuthal flow $v_\varphi$
for $Re=1600$ for shorter heights: $H=8\text{cm}$, $6\text{cm}$, 
$2\text{cm}$, and $1\text{cm}$.
The corresponding picture for 
$H=10\text{cm}$ is Fig.~\ref{fig:vertical_combined_vrz}(e).
}
\end{center}\end{figure}
The numerical simulations presented so far have been performed
for cylinders of height $H=10\text{cm}$, hence about the same as
the width of the gap ($r_2-r_1=11.1\text{cm}$), as in
our laboratory experiment.
In order to elucidate the effects of the
top and bottom endcaps on the fluid motion,
we have also performed numerical experiments
for shorter heights:
$H=8, 6, 2,$ and $1\text{cm}$.
Figure~\ref{fig:short_heights} shows the corresponding flow
profiles after nonlinear saturation.
The Reynolds number, which is based on the cylinder radii
rather than $H$ [eq.(\ref{eq:Reynolds})] is $1600$ in all cases.
Compare these with Fig.~\ref{fig:vertical_combined_vp}(e).

In the case of the shortest height ($H=1\text{cm}$), most of the
fluid rotates at the angular velocity of the endcaps.  This is not
surprising, since in the limit of infinitesimal height, the fluid
would ``adhere'' to the endcaps.  The more rapidly rotating inner
cylinder influences the flow over a radial distance comparable to $H$.
Comparing the panels of Fig.~\ref{fig:short_heights}, we see that the
domain of rigid rotation shrinks as $H$ grows, but it still exists
when $H=10\text{cm}$, as summarized in the schematic
Fig.~\ref{fig:schematic_flow_structure}(b).  The existence of such a
region in a short Couette flow has also been reported by
Dunst\cite{dunst:1972}.

\subsection{Spin Down}\label{subsec:spindown}

\begin{figure}\begin{center}
\includegraphics[width=0.5\textwidth]{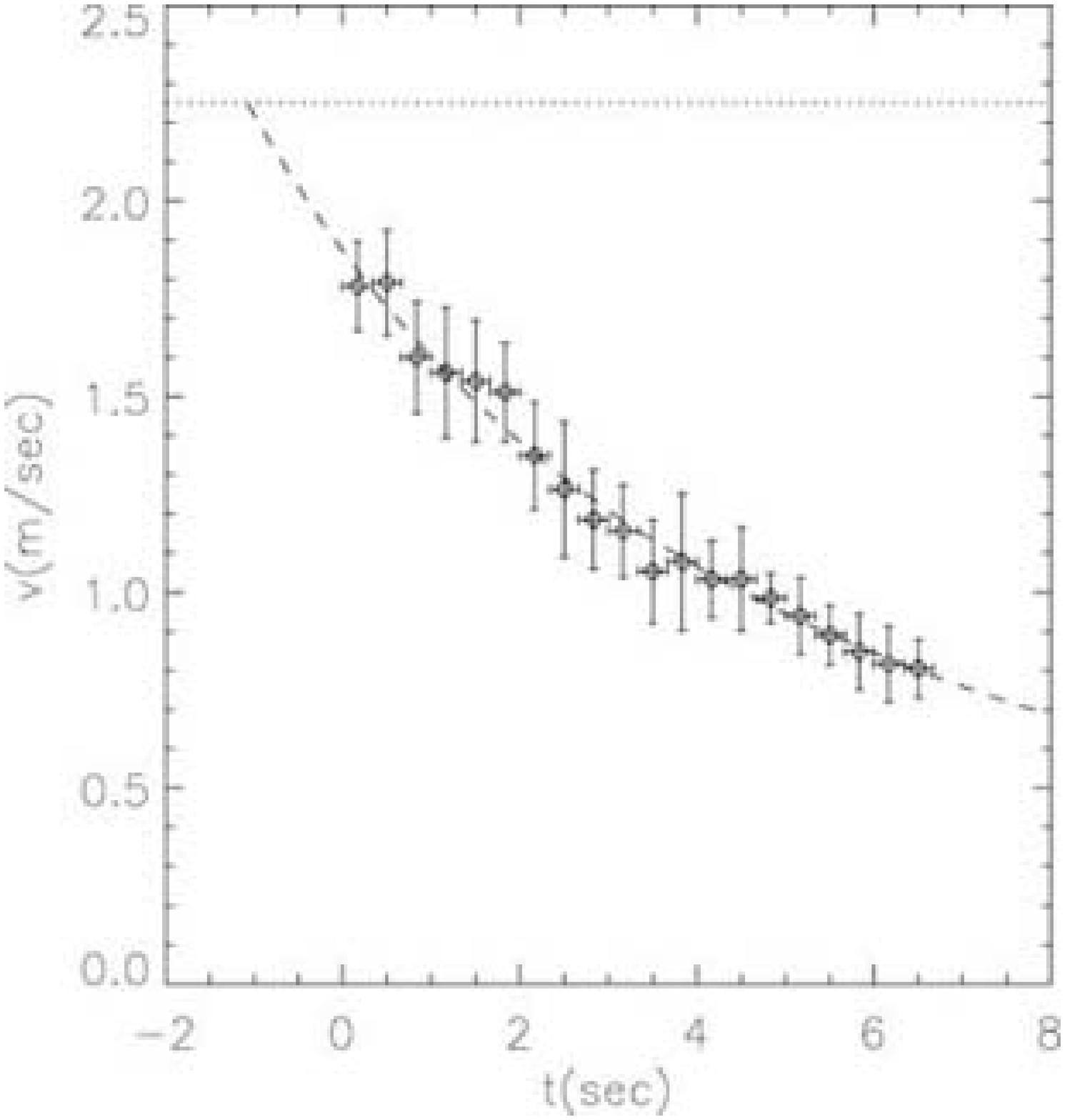}
\caption{\label{fig:spindown_exp}
Experimentally measured azimuthal velocity in the region
$(r,z)\in(11.5\pm 1,\,3 \pm 0.5)\,\text{cm}$
during spin-down after both cylinders and endcaps are stopped
at approximately $t=-1$ sec.
Dotted line is azimuthal velocity in
steady state when $\Omega_1$ = 2000 rpm and $\Omega_2$= 150 rpm.
Dashed line is a fit of the form (\ref{eq:spindown}).}
\end{center}\end{figure}

To better quantify the effects of circulation,
a series of experiments and simulations have been performed
to study the transient flow
when both cylinders (and endcaps) are suddenly stopped.
The rate of spin down reflects the efficiency with which
the circulation transports angular momentum and the viscous
coupling to the walls.
In the experiments,  starting from steady rotation
both cylinders were braked to a complete stop within about one second.
The flow speed in a small volume was measured against time, as shown in
Fig.~\ref{fig:spindown_exp}.

A simple exponential fit to the measured data is not appropriate
because the spin-down time $\tau$ itself depends on angular velocity:
\begin{equation}
\tau = \frac{H}{2 \delta_E \bar \Omega}
= \frac{H}{2 \sqrt{\nu \bar \Omega}},
\label{eq:tau}
\end{equation}
where the factor 2 comes from the fact that the circulation has 2
cells and the Ekman layer thickness $\delta_E$ is taken to be
$\sqrt{\nu/\bar \Omega}$ ($\bar \Omega$ is an averaged angular 
velocity). Thus, we have
\begin{equation}
\frac{d \bar \Omega}{d t} = - \frac{\bar \Omega}{\tau}
\propto - \bar \Omega^{3/2},
\end{equation}
which leads to
\begin{equation}
\bar\Omega(t) = \frac{\bar \Omega (t_0)}
{\left( 1 + \frac{t - t_0}{\tau} \right)^2}.
\label{eq:spindown}
\end{equation}
The measurements are fitted to eq.(\ref{eq:spindown})
where the steady state angular velocity 
$\bar\Omega(t_0)$ is known while $t_0$ and $\tau$ are fitting 
parameters. The fitted line is shown by the dashed line in 
Fig.~\ref{fig:spindown_exp} and 
the spin down time, $\tau=11.2\pm 0.9$ sec, is obtained.

\begin{figure}\begin{center}
\includegraphics[width=0.5\textwidth]{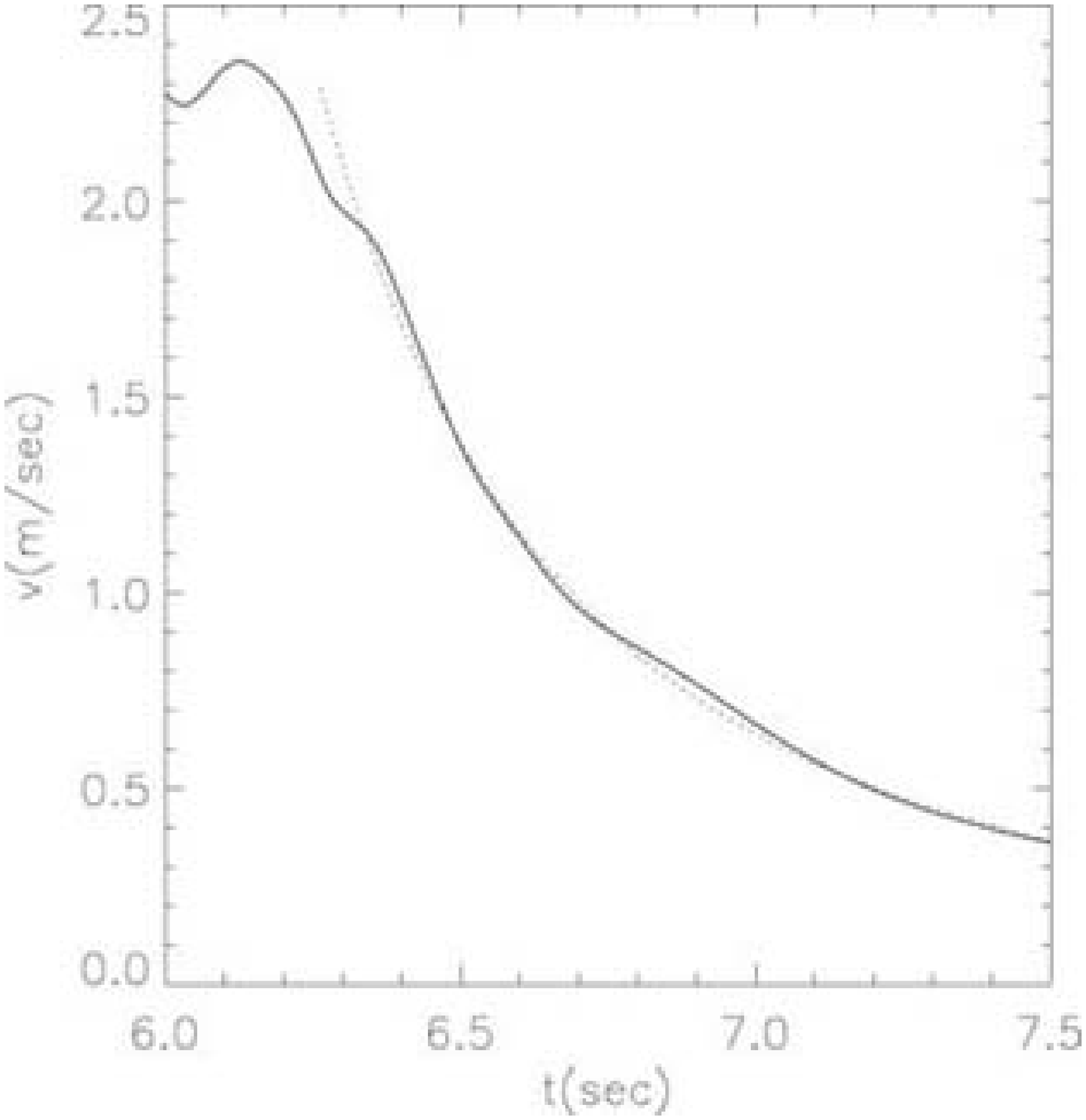}
\caption{\label{fig:spindown_sim}
Like Fig.~\ref{fig:spindown_exp}, but for a simulation at initial
$Re=3200$.
The boundaries are stopped at $t=6.0747\text{ sec}$.}
\end{center}\end{figure}

\begin{figure}\begin{center}
\includegraphics[width=0.5\textwidth]{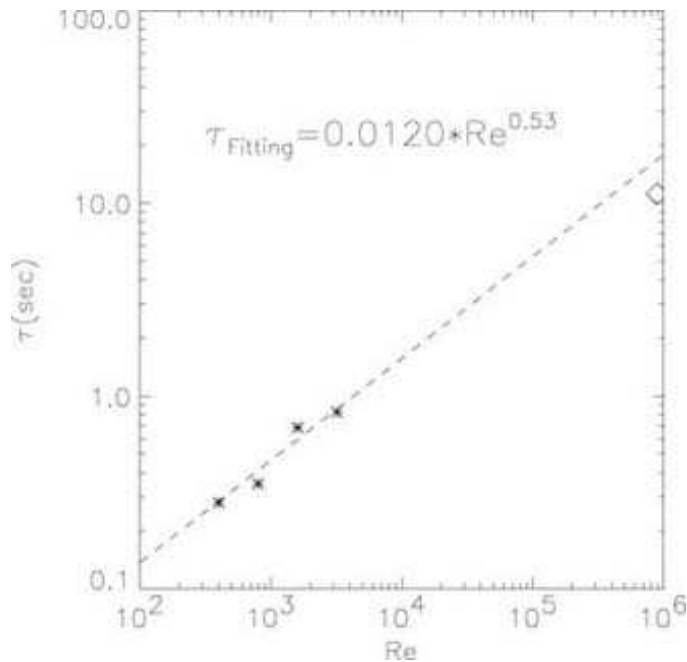}
\caption{\label{fig:spindown_scaling}
Spin down time versus $Re$ for simulations (asterisks) and experiment (diamond).
Dotted line is a fitted curve to the results from simulation only.}
\end{center}\end{figure}

Spin down has also been simulated by imposing a sudden stop
of all boundaries. Figure \ref{fig:spindown_sim} shows results
for azimuthal velocity in the same volume as in the experiment
for $Re=3200$. Again, the spin down time is obtained
by fitting eq.(\ref{eq:spindown}), with the result
$\tau = 0.82$ sec. Figure \ref{fig:spindown_scaling} displays
$\tau$ determined similarly for a series of simulations at different
$Re$.  The trend is well fit by a power law (dashed line),
\begin{equation}\label{eq:taufit}
    \tau = 0.012Re^{0.53}\,\text{sec}.
\end{equation}
For the purposes of Fig.~\ref{fig:spindown_scaling} and eq.(\ref{eq:taufit}),
we fix $\Omega_1=2000\,\text{rpm}$  and $\Omega_2=150\,\text{rpm}$ (as in
the experiment) and imagine that the Reynolds number of the simulations
is controlled by varying the viscosity.  If the viscosity is
fixed and the rotation rates vary (as would be more convenient in
an experiment) then the spindown time scales as $Re^{-0.47}$.
The power law (\ref{eq:taufit}) agrees excellently with the simple estimate
given by eq.(\ref{eq:tau}),
\begin{equation}
    \tau=\frac{H}{2 \sqrt{\nu \bar \Omega}}
        =\frac{H}{2 
    \sqrt{r_1(r_2-r_1)\Omega_1 \bar \Omega\,Re^{-1}}}
    =0.011 Re^{1/2}\,\text{sec}.
    \label{eq:scaling}
\end{equation}
[For comparison, eq.(\ref{eq:Ekmantime}) predicts 
$t_E=0.0095 Re^{1/2}\,\text{sec}$.]
The experimental point, which was not included in the fit, is rather
close to the extrapolation of eq.(\ref{eq:taufit}): $11.2\pm0.9\,\text{sec}$
observed \emph{vs.} $17\pm0.9\,\text{sec}$ predicted, using
$\nu=0.01\,{\rm cm^2\,sec^{-1}}$ for water.

\section{Summary and Discussion} \label{section:summary}
Motivated by our proposed magnetohydrodynamic
experiment\cite{ji:2001,goodman:2002},
we have studied a short, wide-gap, hydrodynamic Couette flow
by experiment and numerical simulation.
A challenge of the gallium experiment will be to 
set up an initial rotation profile
that is stable to the Taylor-Couette instability (TCI) while
unstable to the magnetorotational instability (MRI)
when an appropriate magnetic field is applied.
It is found experimentally that the
azimuthal flow $v_\varphi(r)$ is significantly different from that in
an infinitely long circular Couette system.
In order to understand the underlying physics, 
numerical simulations have been performed
using the same boundary conditions as in the experiment.
The measured profile of $v_\varphi$ is successfully 
reproduced by the simulations, which show
a strong poloidal circulation driven by boundary layers at the endcaps.
Furthermore, excellent agreement between 
experiment and simulation is found for
the spin-down time when the Reynolds number is scaled.

These agreements are rather remarkable considering that there is a
factor of 300 between the Reynolds numbers of the simulations and of
the experiments, and that the simulations are performed in two
dimensions, while the experiments are three-dimensional.  The
suggestion is that the essential dynamics of the system do not change
when $Re$ is raised from 3200 to about $10^6$.

Is it plausible that the the boundary layer remains laminar up to the
experimental Reynolds number $Re= 8.8\times 10^5$?  Perhaps:
nonrotating flow over a flat plate remains laminar below $Re_{\rm
crit}\approx 5\times 10^5$; perhaps more relevantly for our
experiment, $Re_{\rm crit}\approx 3\times 10^5$ for a disk spinning
freely in an extended, nonrotating fluid, where $Re\equiv\Omega
R^2/\nu$ based on the angular velocity and radius of 
the disk\cite{Schlichting:1979}.  For fully turbulent boundary layers, the
stress ($\sigma=$ lateral force per unit area) exerted on the boundary
is parametrized by a \textit{friction coefficient} 
$C_f\equiv \left.\sigma\right/\frac{1}{2}\rho V_\infty^2$,
where $V_\infty$ is the relative velocity
of the fluid well outside the boundary layer.
It is known that $C_f$ varies slowly with $Re$ when $Re$ is large.
In flow over smooth surfaces at $Re\gg 10^6$, for example,
von K\'arman prescribes\cite{Schlichting:1979}
$C_f=0.455(\log_{10} Re)^{-2.58}$.  Let us
suppose that we can take $C_f$ to be constant over the entire
boundary (cylinders and endcaps) for the purpose of estimating
the spin-down rate.  Taking advantage of the fact that the
specific angular momentum varies slowly within our
steady-state Couette flow and defining
$\bar J\equiv(r_1^2\Omega_1^2+r_2^2\Omega_2^2)/2$, we estimate
that the total torque on the fluid shortly after the cylinders stop
is $\Gamma\approx2\pi\rho(H+r_2-r_1)\bar J^2 C_f$.
The total angular
momentum of the fluid is $L\approx\pi\rho(r_2^2-r_1^2)H\bar J$.
Thus the spin-down time becomes
\[
\tau_f\equiv\frac{L}{\Gamma}=\frac{H(r_2^2-r_1^2)}
{(r_2^2\Omega_2+r_1^2\Omega_1)(H+r_2-r_1)}\approx 0.015 C_f^{-1}
\,\text{sec}.
\]
This agrees with the observed value of $11\,\text{sec}$ for 
$C_f=1.3\times 10^{-3}$.
For comparison, von K\'arman's formula
predicts $C_f(10^6) \approx 4.5\times10^{-3}$.  Our Reynolds number
however, is not far from $Re_{\rm crit}$, so turbulence may not
be fully developed.  Indeed, alternative definitions of the
Reynolds number fall even closer to the critical value:
for example, $\bar J/\nu\approx 3.0\times10^5$.


Detailed analysis of the simulations show that the poloidal
circulation consists of two cells.  A strong radially inward flow
forms near each of the endcaps in a thin boundary layer. After
turning into a vertical flow along the inner cylindrical wall, these
layers merge at the midplane into a jet-like, radially outward flow to
complete the circulation. The existence of such a jet-like feature
appears not to have been recognized previously.  Dunst performed a
water experiment in short cylindrical annulus with similar condition
to ours\cite{dunst:1972}.  One set of five experiments by Dunst was
carried out with rigid endcaps fixed to the outer cylinder.  (Other
experiments were done with a free upper surface.)  Dunst reported the
formation of a two-cell pattern as well as a region of rigid-body
rotation in the outer part of the fluid.  However, there was no
description of a jet between the cells.  It is possible that 
the jet was just overlooked.  We
note that although we have preliminary evidence for a jet-like flow at
the midplane between two cells in the experiment, it is difficult to
visualize it and to measure its detailed characteristics.  We also
carried out simulations in which the outer cylinder is stationary,
leading presumably to TCI instabilities.  It is interesting that the
jet does not form in the latter situation, which has been the main
focus of experimental effort in short Couette flows.

We have seen that a region of rigid-body rotation occurs in the outer
part of the system.  This has been explained by the tendency of the
flow to ``adhere'' to the outer cylinder and endcaps.  An alternative
setup of the apparatus would be to have the caps rotate rigidly with
the inner rather than the outer cylinder.  In that case, rigid-body
rotation would be expected to appear in the inner part of the flow.
Actually, such a fluid-dynamical system has been investigated in the
literature in connection with hard disk drives for
computers\cite{abrahamson:1989,humphrey:1995,kaneko:1977,lennemann:1974,schuler:1990}.
It has been shown in these studies that most of the fluid rotates
rigidly with the disks.

The poloidal circulation, and especially the jet, found in this study
are an interesting phenomenon in rotating fluids.
They transport angular momentum efficiently and reduce the
free energy available for shear-driven instabilities.
Therefore, we will need to minimize this circulation in the MRI
experiment.

\begin{figure}\begin{center}
\includegraphics[width=0.45\textwidth]{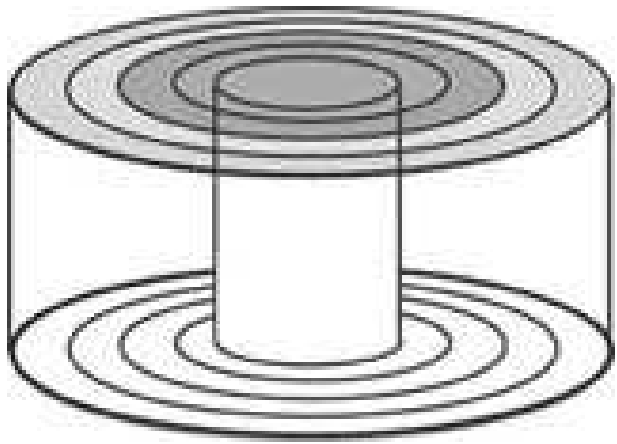}
\caption{\label{fig:system_multiring}Experimental
setup with independently rotating rings in the endcaps.
}
\end{center}\end{figure}

\begin{figure}\begin{center}
\includegraphics[width=0.40\textwidth]{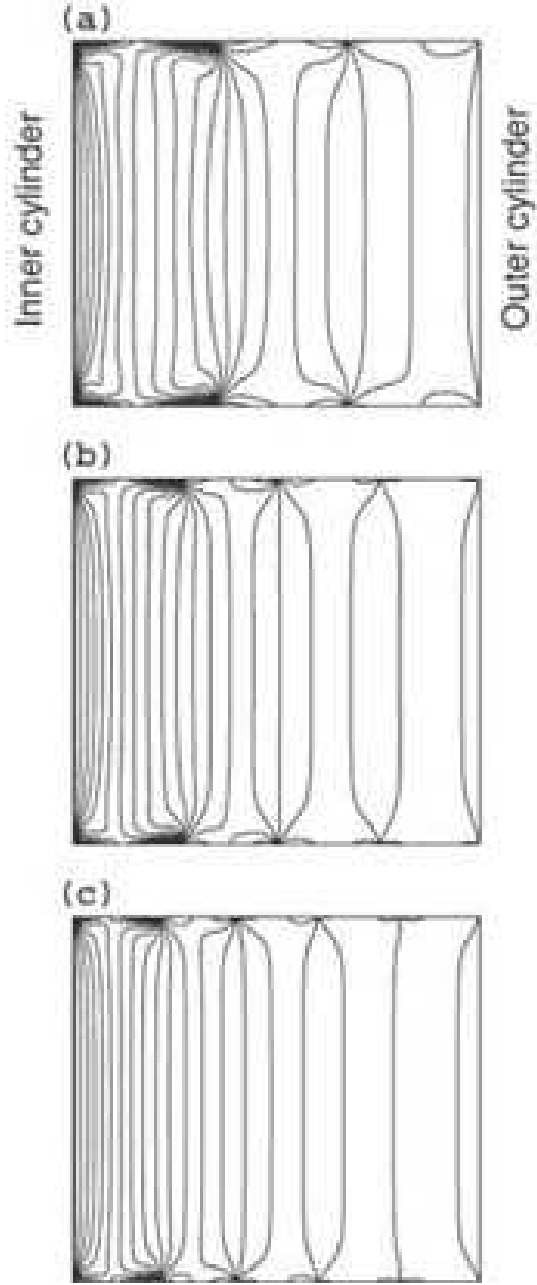}
\caption{\label{fig:vp_multirings}Simulated azimuthal
velocity $v_\varphi$ for $Re=1600$ when the endcaps
are divided into multiple rings;
(a) 3 rings;
(b) 4 rings;
and (c) 5 rings.
}
\end{center}\end{figure}

\begin{figure}\begin{center}
\includegraphics[width=0.45\textwidth]{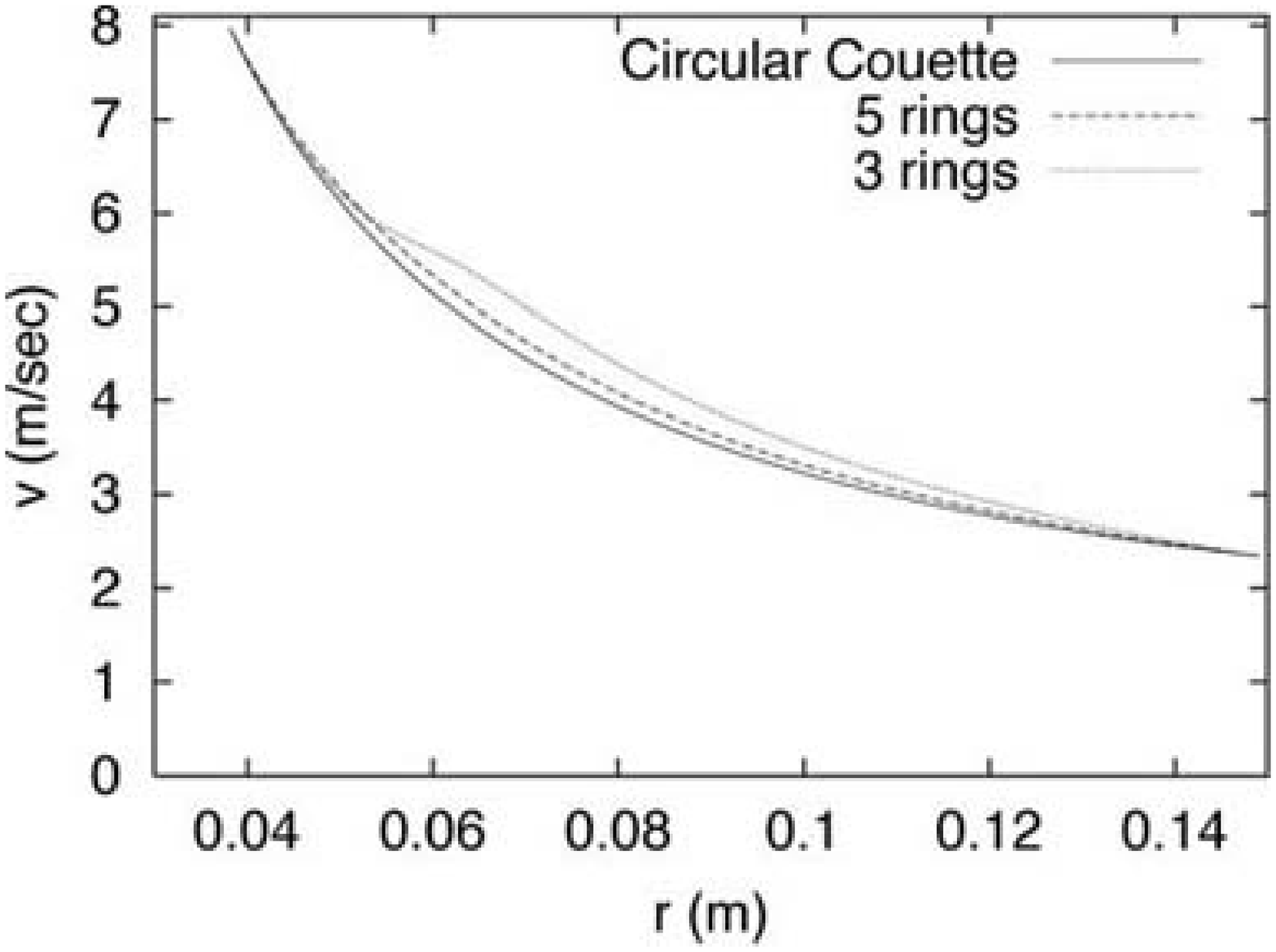}
\caption{\label{fig:vp_graph_multiring}Radial profile
of $v_\varphi$ at the middle height 
when the dividing number of the cap is $3$ and $5$.
See also Fig.~\ref{fig:vp_multirings}(b) and (c).
}
\end{center}\end{figure}

One idea for reducing the effect of endcaps 
is to use a tapered section\cite{Cannell:1983}.
Another idea is to divide the cap into two parts:
the inner one fixed to the inner cylinder,
and the outer one fixed to the outer cylinder.
Here, we expand on the latter idea.
We divide the endcaps into multiple rings
that rotate independently
(see Fig.~\ref{fig:system_multiring}).
The angular velocity of each ring is chosen according to
its center radius and the circular Couette flow.

Figure~\ref{fig:vp_multirings} shows the
profiles of $v_\varphi$ at $Re=1600$ for three
choices of the number of rings.
The parallel contour lines in the figure
indicate that the fluid is in a Taylor-Proudman state.
The effect of the endcaps is highly localized in its vicinity
and the poloidal circulation is suppressed almost perfectly.
Without the circulation, the stationary flow in this
differentially rotating system is very close
to that of an infinitely long, ideal circular Couette flow.
This is confirmed by 
the $v_\varphi$ profile at the midplane
when 3 and 5 rings are used (Fig.~\ref{fig:vp_graph_multiring}).

The rings will reduce the relative velocity between
the boundary and the interior flow in steady state, to the point
where a turbulent boundary is unlikely (see the discussion above).
This makes it more likely that the simulations accurately predict
the interior flow despite their relatively low Reynolds number.

An optimization process incorporating this idea is underway to design
a gallium experiment with maximum controllability of the flow profile,
yet with minimum complications to engineering and experimental
operations.

\section*{Acknowledgments}
This work was carried out when
one of the authors (A.K.) was staying
at Princeton Plasma Physics Laboratory.
A.K. thanks to W.~Tang for his hospitality.
Financial support to A.K. by
Japanese Ministry of Education, Culture, Sports, and Technology
is gratefully acknowledged.
We used NEC SX-5 of
National Institute for Fusion Science, Japan
for the numerical simulations.
The experimental work was supported by the US Department
of Energy.  Support has also been received from the National
Science foundation under grant AST-0205903.
Technical assistance from R. Cutler
is gratefully acknowledged.

\appendix
\section{Ekman layer in a differentially rotating flow}

Consider a steady axisymmetric
flow that departs only
slightly from a centrifugally stable Couette profile,
\begin{equation}\label{eq:couetteprof}
\Omega=a+\frac{b}{r^2},\qquad a,b>0 \mbox{ \& constant}.
\end{equation}
One may linearize the equations of motion
(\ref{eq:sim_04})-(\ref{eq:sim_13}) about the basic state
(\ref{eq:couetteprof}).  Marking first-order quantities with the
prefix $\delta$, we have $\psi\to\delta\psi$,
$\omega_\varphi\to\delta\omega_\varphi$, and $v_\varphi\to
r\Omega+\delta v_\varphi$.  Linearizing eq.(\ref{eq:sim_04}) and
discarding time derivatives,
\[
2\Omega\partial_z\delta v_\varphi + \nu\Delta\omega_\varphi=0,
\]
where $\Delta\equiv(\nabla^2-r^{-2})$.  Incompressibility
implies $\partial_z\delta v_z= -r^{-1}\partial_r(r\delta v_r)$, whence
\[
\partial_z\delta\omega_\varphi=\partial_z\left(\partial_z\delta v_r
-\partial_r\delta v_z\right)=\Delta\delta v_r\,,
\]
so that
\begin{equation}\label{eq:radbal}
2\Omega\partial_z^2\delta v_\varphi = -\nu\Delta^2\delta v_r\,.
\end{equation}
By similar steps, eq.(\ref{eq:sim_05}) yields
\begin{equation}\label{eq:azibal}
\frac{1}{r}\frac{dJ}{dr}\delta v_r= \nu\Delta v_\varphi\,.
\end{equation}
For the Couette profile (\ref{eq:couetteprof}), the coefficient of $\delta v_r$
in this last equation is $2a$, a constant.  Eliminating $\delta v_r$ between
eqs.(\ref{eq:radbal})\&(\ref{eq:azibal}) yields
\begin{equation}\label{eq:dvphieqn}
\left(\kappa^2\partial_z^2+\nu^2\Delta^3\right)\delta v_\varphi=0\,.
\end{equation}
When viscosity can be neglected and $\kappa^2>0$,
eqs.(\ref{eq:radbal})-(\ref{eq:dvphieqn}) imply that small, low-frequency 
($\ll\kappa$) motions are independent of $z$: this is the Proudman theorem.

We apply these equations to the boundary layer at one of the
endcaps, which rotates differentially  (as might be
achieved by dividing it into many rings, see \S\ref{section:summary})
with a slightly different profile $\Omega_{\rm cap}(r)$ from that of 
the fluid in the interior, eq.(\ref{eq:couetteprof}).
Vertical derivatives are much larger than radial ones in the boundary
layer, so eq.(\ref{eq:dvphieqn}) implies that the boundary-layer behavior
is $\delta v_\varphi\propto\exp(kz)$ with 
\begin{equation}\label{eq:kroots}
k^4=-(\kappa/\nu)^2,\quad
k=\left(\pm 1\pm i\right)\sqrt{\frac{\kappa}{2\nu}}\,.
\end{equation}
Of these four roots, only the two for which $\delta v_\varphi$
decays with distance from the boundary are admissible.  To be definite,
let us consider the lower endcap, so that $\Re(k)<0$.  The Ekman
layer thickness is
\begin{equation}\label{eq:kroot}
\delta_E\equiv\sqrt{\frac{2\nu}{\kappa}}\,,
\end{equation}
and
\begin{eqnarray}
\delta v_\varphi&=& r\left(\Omega_{\rm cap}-\Omega\right)e^{-z/\delta_E}
\cos(z/\delta_E),\nonumber\\
\delta v_r&=& \frac{\kappa}{a}
\left(\Omega_{\rm cap}-\Omega\right)e^{-z/\delta_E}\sin(z/\delta_E)
\label{eq:dvrsoln}
\end{eqnarray}
if we take $z=0$ at the endcap rather than the midline of the cylinders.
The radial mass flow is
\begin{equation}\label{eq:MdotE}
\dot M_E=~2\pi\rho r\int\limits_0^\infty\delta v_r\,dz~=~
\pi\rho r^2\left(\Omega_{\rm cap}-\Omega\right)\frac{\kappa}{a}\,\delta_E.
\end{equation}
The net torque exerted on the fluid by both endcaps is
\[
\Gamma = 2\int\limits_{r_1}^{r_2}\dot M_E\frac{dJ}{dr}\,dr
~=~4\pi\rho\int\limits_{r_1}^{r_2} \kappa\delta_E\, 
r^3(\Omega_{\rm cap}-\Omega)\,dr\,,
\]
and the amount of angular momentum that must be added to the fluid
to make its rotation profile agree with the endcaps is
\[
L'-L=2\pi\rho H\int\limits_{r_1}^{r_2} r^3(\Omega_{\rm cap}-\Omega)\,dr\,.
\]
We may estimate the spin-up or spin-down time as
\begin{equation}\label{eq:Ekmantime}
t_E=~\frac{L'-L}{\Gamma}
~\approx~\frac{H}{\sqrt{8\nu\bar\kappa}}\,.
\end{equation}
Here $\bar\kappa$ is a weighted average over radius; if we approximate
it by eq.(\ref{eq:kappabar}) then eq.(\ref{eq:Ekmantime})
predicts $t_E\approx 9\,\mbox{sec}$ for water with
the values of $r_1$, $r_2$, $\Omega_1$, $\Omega_2$, and $H$ in
Figs.~(\ref{fig:apparatus}) \& (\ref{fig:spindown_exp}).  The agreement
with the measured value ($11.2\pm0.9~\mbox{sec}$) is perhaps better
than we deserve in view of the crudeness of the theoretical treatment.
In particular, since $\Omega_{\rm cap}=0$ in the spindown experiment,
our linear approximation is not applicable.


\begin{thebibliography}{10}

\bibitem{lundquist:1949a}
S. Lundquist.
Experimental demonstration of magnetohydrodynamic waves.
{\em Nature}, 164:145--146, 1949.

\bibitem{nakagawa:1955}
Y. Nakagawa.
An experiment on the inhibition of thermal convection by a magnetic
  field.
{\em Nature}, 175:417--419, 1955.

\bibitem{gailitis:2000}
A.~Gailitis, O.~Lielausis, S.~Dement'ev, E.~Platacis, A.~Cifersons, G.~Gerbeth,
  T.~Gundrum, F.~Stefani, M.~Christen, H.~Haenel, and G.~Will.
Detection of a flow induced magnetic field eigenmode in the \mbox{Riga}
  dynamo facility.
{\em Phys. Rev. Lett.}, 84:4365--4368, 2000.

\bibitem{ji:2001}
H. Ji, J. Goodman, and A. Kageyama.
Magnetorotational instability in a rotating liquid metal annulus.
{\em Mon. Not. Astron. Soc.}, 325:L1--L5, 2001.

\bibitem{goodman:2002}
J. Goodman and H. Ji.
Magnetorotational instability of dissipative couette flow.
{\em J. Fluid Mech.}, 462:365--382, 2002.

\bibitem{taylor:1923}
G.~I. Taylor.
Stability of a viscous liquid contained between two rotating
  cylinders.
{\em Phil. Trans. Roy. Soc. London A}, 223:289--343, 1923.

\bibitem{benjamin:1978a}
T.~B. Benjamin.
Bifurcation phenomena in steady flows of a viscous fluid. {I}.
  theory.
{\em Proc. Roy. Soc. London A}, 359:1--26, 1978.

\bibitem{benjamin:1978b}
T.~B. Benjamin.
Bifurcation phenomena in steady flows of a viscous fluid. {II}.
  experiments.
{\em Proc. Roy. Soc. London A}, 359:27--43, 1978.

\bibitem{benjamin:1981}
T.~B. Benjamin and T.~Mullin.
Anomalous modes in the {T}aylor experiment.
{\em Proc. Roy. Soc. London A}, 377:221--249, 1981.

\bibitem{hall:1982}
P. Hall.
Centrifugal instabilities of circumferential flows in finite cylinders: the wide gap problem.
{\em Proc. Roy. Soc. London A}, 384:359--379, 1982.

\bibitem{lucke:1984}
M. L\"ucke and M. Mihlcic and K. Wingerath.
Flow in small annulus between concentric cylinders.
{\em J. Fluid Mech.}, 140:343--353, 1984.

\bibitem{aitta:1985}
A. Aitta and Guenter Ahlers and David S. Cannel.
Tricritical Phenomena in Rotating Couette-Taylor Flow.
{\em Phys. Rev. Lett.}, 54:673--676, 1985.

\bibitem{heinrichs:1986}
R. Heinrichs and G. Ahlers and D.S.~Cannell.
Effects of Finite Geometry on the Wave Number of Taylor-Vortex Flow.
{\em Phys. Rev. Lett.}, 56:1794--1797, 1986

\bibitem{pfister:1988}
G.~Pfister and H.~Schmidt and K.~A.~Cliffe and T.~Mullin.
Bifurcation phenomena in Taylor--Couette flow in a very short annulus.
{\em J. Fluid Mech.}, 191:1--18, 1988.

\bibitem{tavener:1991}
S.J. Tavener and T. Mullin and K.A. Cliffe.
Novel bifurcation phenomena in a rotating annulus.
{\em J. Fluid Mech.}, 229:483--497, 1991.

\bibitem{cliffe:1992}
K.~A.~Cliffe and J.~J.~Kobine and T.~Mullin.
The Role of Anomalous Modes in Taylor--Couette Flow.
{\em Int. J. Heat Mass Transfer}, 439:341--357, 1992


\bibitem{sobolik:2000}
V.~Sobol\'{\i}k and B.~Izrar and F.~Lusseyran and S.~Skali.
Iteraction between the Ekman layer and the Couette-Taylor instability.
{\em Int. J. Heat Mass Transfer}, 43:4381--4393, 2000

\bibitem{furukawa:2002}
H.~Furusawa and T.~Watanabe and Y.~Toya and I.~Nakamura.
Flow pattern exchange in the Taylor-Couette system with a very small aspect ratio.
{\em Phys. Rev. E}, 65:036306--036312, 2002.

\bibitem{mullin:2002}
T. Mullin and Y. Toya and S.J. Tavener.
Symmetry breaking and multiplicity of states in small aspect ratio Taylo--Couette flow.
{\em Phys. Fluids}, 14:2778--2787, 2002.

\bibitem{lopez:2003}
J.M. Lopez and F. Marques.
Small aspect ratio Taylor--Couette  flow: Onset of a very-low-frequency three-torus state.
{\em Phys. Rev. E}, 68:036302-1--9, 2003

\bibitem{schulz:2003}
A. Schulz and G. Pfister and S.J. Tavener,
The effect of outer cylinder rotation on Taylor--Couette flow at small aspect ratio.
{\em Phys. Fluids}, 15:417--425, 2003


\bibitem{dunst:1972}
M.~Dunst.
An experimental and analytical investigation of angular momentum
  exchange in a rotating fluid.
{\em J. Fluid Mech.}, 55:301--310, 1972.

\bibitem{batchelor:1967}
G.K. Batchelor.
{\em An Introduction to Fluid Dynamics}.
Cambridge University Press, 1967.

\bibitem{noguchi:2002}
K.~Noguchi, V.I. Pariev, S.A. Colgate, H.F. Beckley, and J.~Nordhaus.
Magnetorotational instability in liquid metal couette flow.
{\em Astrophys. J.}, 575:1151--1162, 2002.

\bibitem{ferziger:2002}
J.~H. Ferziger and M.~Peri\/{c}.
{\em Computational Methods for Fluid Dynamics}.
Springer, 3rd edition, 2002.

\bibitem{mullin:2000}
T. Mullin, D. Satchwell, and Y. Toya.
Pitchfork bifurcation in small aspect ratio taylor--couette flow.
In Christoph Egbers and Gerd Pfister, editors, {\em Physics of
  Rotating Fluids}, pages 3--21. Springer, 2000.

\bibitem{wendl:1999}
M.C. Wendl.
General solution for the couette flow profile.
{\em Phys. Rev. E}, 60:6192--6194, 1999.


\bibitem{chandrasekhar:1968}
S.~Chandrasekhar.
{\em Hydrodynamic and Hydromagnetic Stability}.
Dover, 1968.

\bibitem{Schlichting:1979}
H.~Schlichting.
{\em Boundary-{L}ayer {T}heory}.
McGraw-Hill, New York, 7 edition, 1979.

\bibitem{abrahamson:1989}
S.D. Abrahamson, J.K. Eaton, and D.J. Koga.
The flow between shrouded corotating disks.
{\em Phys. Fluids A}, 1:241--251, 1989.

\bibitem{humphrey:1995}
J.~A.~C. Humphrey, C.~A. Schuler, and D.~R. Webster.
Unsteady laminar flow between a pair of disks corotating in a fixed
  cylindrical enclosure.
{\em Phys. Fluids}, 7:1225--1240, 1995.

\bibitem{kaneko:1977}
R. Kaneko, S. Oguchi, and K. Hoshiya.
Hydrodynamic characteristics in disk packs for magnetic storage.
{\em Review of the Electrical Communication Laboratories, Nippon
  Telegraph and Telephone Public Corp., Japan}, 25:1325--1336, 1977.

\bibitem{lennemann:1974}
E.~Lennemann.
Aerodynamic aspects of disk files.
{\em IBM J. Res. Develop.}, pages 480--488, November 1974.

\bibitem{schuler:1990}
C.~A. Schuler, W.~Usry, B.~Weber, J.~A.~C. Humphrey, and R.~Greif.
On the flow in the unobstructed space between schrouded corotating
  disks.
{\em Phys. Fluids A}, 2:1760--1770, 1990.

\bibitem{Cannell:1983}
D.S. Cannell, M.A. Dominguez-Lerma, and G. Ahlers.
Experiments on wave number selection in rotating couette-taylor-flow.
{\em Phys. Rev. Lett.}, 50:1365--1368, 1983.

\end{thebibliography}


\end{document}